\documentclass[10pt]{article}

\usepackage{fullpage,graphicx,subfigure,mathdots,mathpazo}
\usepackage{amsmath,amscd,tikz,mathrsfs}
\usepackage[normalem]{ulem}
\usepackage{amsmath}
\usepackage{setspace}

\usepackage{epsfig,amsmath,amsthm,graphicx,amssymb,overpic}
\usepackage{epsfig,amsmath,graphicx,amssymb,overpic,cite}
\usepackage{listings,diagbox}
\usepackage{epsfig,amsmath,graphicx,amssymb,overpic}
\usepackage{bm}
\usepackage{graphicx}
\usepackage{subfigure}

\usepackage{pdfsync}
\usepackage{cite}
\usepackage{amscd}
\usepackage{amsmath}
\usepackage{latexsym}
\usepackage{amsfonts}
\usepackage{amssymb}
\usepackage{amsthm}
\usepackage{graphicx}
\usepackage{indentfirst}
\usepackage{enumerate}
\usepackage{amstext}
\usepackage[
            pdfstartview=FitH,
            CJKbookmarks=true,
            bookmarksnumbered=true,
            bookmarksopen=true,
            pdfborder=001,   
            linkcolor=blue,
            ]{hyperref}

 \allowdisplaybreaks[4]

\def\no{\nonumber}
\def\be{\begin{equation}}
\def\ee{\end{equation}}
\def\bee{\begin{eqnarray}}
\def\ene{\end{eqnarray}}
\def\bes{\begin{subequations}}
\def\ees{\end{subequations}}

\newcommand{\PT}{\mathcal{PT}}

\def\v{\vspace{0.1in}}
\def\d{\displaystyle}

\begin{document}

\baselineskip=13pt
\renewcommand {\thefootnote}{\dag}
\renewcommand {\thefootnote}{\ddag}
\renewcommand {\thefootnote}{ }

\pagestyle{plain}

\begin{center}
\baselineskip=16pt \leftline{} \vspace{-.3in} {\Large \bf  Higher-dimensional soliton generation, stability and excitations of the $\mathcal{PT}$-symmetric nonlinear Schr\"odinger equations} \\[0.1in]
\end{center}


\begin{center}
Yong Chen$^{1}$,\, Zhenya Yan$^{2,3,*}$\footnote{$^{*}${\it Email address}: zyyan@mmrc.iss.ac.cn (Corresponding author)},\,
 Boris A. Malomed$^{4,5}$  \\[0.05in]
{\it $^1$ School of Mathematics and Statistics, Jiangsu Normal
University, Xuzhou 221116, China \\
$^2$Key Lab of Mathematics Mechanization, Academy of Mathematics and Systems
Science, \\ Chinese Academy of Sciences, Beijing 100190, China\\
$^3$School of Mathematical Sciences, University of Chinese Academy of
Sciences, Beijing 100049, China\\
$^4$Department of Physical Electronics, School of Electrical Engineering,
Faculty of Engineering, \\ and Center for Light-Matter Interaction, Tel Aviv
University, Tel Aviv 69978, Israel\\
$^5$Instituto de Alta Investigaci\'{o}n, Universidad de Tarapac\'{a},
Casilla 7D, Arica, Chile} \\
\end{center}

\baselineskip=13pt

\baselineskip=13pt

\vspace{0.1in}
 {\bf Abstract.} We study a class of physically intriguing $\mathcal{PT}$-symmetric generalized
Scarf-II (GS-II) potentials, which can support exact solitons in one- and multi-dimensional nonlinear Schr\"odinger equation.
In the 1D and multi-D settings, we find that a properly adjusted localization parameter may support fully real energy spectra.
Also, continuous families of fundamental and higher-order solitons are produced.
The fundamental states are shown to be stable, while the higher-order ones, including 1D multimodal solitons,
2D solitons, and 3D light bullets, are unstable. Further, we find that the
stable solitons can always propagate, in a robust form, remaining trapped in
slowly moving potential wells of the GS-II type, which opens the way for
manipulations of optical solitons. Solitons may also be transformed into
stable forms by means of adiabatic variation of potential parameters.
Finally, an alternative type of $n$-dimensional $\mathcal{PT}$-symmetric
GS-II potentials is reported too. These results will be useful to further explore the higher-dimensional $\PT$-symmetric
solitons and to design the relative physical experiments.


\vspace{0.1in} {\it Keywords:} Higher-dimensional nonlinear Schr\"odinger equation;
$\mathcal{PT}$-symmetric potentials; stable solitons; adiabatic management





\vspace{0.1in}

\section{Introduction}

In the classical quantum mechanics (QM), a Hamiltonian is usually required to be hermitian,
as the resulting spectra predict physically relevant real eigenvalues of
energy. However, the hermiticity of the Hamiltonian is not a necessary
condition to secure its real spectra. In this respect, the concept of
non-Hermitian $\mathcal{PT}$-symmetric Hamiltonian, first put forward by
Bender and Boettcher in 1998 \cite{bender1998real}, is a noteworthy
generalization of classical QM, where ${\cal P}$ and ${\cal T}$ are defined as
${\mathcal P}:\, x\to -x$ and ${\mathcal T}:\, i\to -i$. Such Hamiltonian include
the complex potentials whose real and imaginary parts are spatially even and
odd, respectively. The $\mathcal{PT}$-symmetric Hamiltonians exhibit fully
real spectra, provided that the strength of the imaginary part of the
potential does not exceed a critical value \cite%
{bender1998real,bagchi2000new,bagchi2001generalized,ahmed2001real,bagchi2002pseudo,bagchi2002non}%
. The natural nonlinear extension of $\mathcal{PT}$-symmetric models produce
diverse stable localized modes ($\mathcal{PT}$-symmetric solitons)~\cite%
{musslimani2008optical,shi2011bright,nixon2012stability,achilleos2012dark,
yan2013complex,lumer2013nonlinearly,yan2015spatial,wen2015dynamical,yan2015solitons}%
. Fundamentally fascinating features associated with the $\mathcal{PT}$
symmetry, including symmetry breaking and presence of exceptional points in
the spectrum, have been investigated theoretically and observed
experimentally in many physical systems \cite%
{guo2009observation,rueter2010observation,regensburger2012parity,castaldi2013pt,regensburger2013observation, peng2014parity,zyablovsky2014pt,chen2016pt,takata2017pt,hodaei2017enhanced,liu2019symmetry}%
.

In particular, the $\mathcal{PT}$ symmetry can be realized in optics by
introducing mutually symmetric gain and loss elements in the waveguide
geometry, which are represented by the above-mentioned spatially odd
imaginary part of the effective potential~\cite%
{ultanir2004dissipative,musslimani2008optical}. In other words, the
respective complex refractive-index distribution may play the role of $%
\mathcal{PT}$-symmetric optical potentials. Particularly, in periodic
photonic-lattice potentials, a large number of new phenomena, maintained by
the $\mathcal{PT}$-symmetry, have been discovered, such as the double
refraction, power oscillations, phase singularities, and secondary emission~%
\cite{makris2008beam,makris2010pt,makris2011pt}.

In the last decade, increasing attention has been drawn to exploring one-
and high-dimensional solitons in nonlinear media with specially designed
effective potentials, including the well-known ones of the Scarf-II type
\cite%
{musslimani2008optical,musslimani2008analytical,yan2013complex,yan2015spatial,dai2014stable}%
, Gaussian \cite{hu2011solitons,achilleos2012dark,yang2014symmetry},
harmonic oscillator~\cite{zezyulin2012nonlinear,dai2014stable}, Rosen-Morse
\cite{midya2013nonlinear}, double-delta~\cite%
{cartarius2012model,single2014coupling}, super-Gaussian~\cite%
{jisha2014influence}, optical lattices and superlattices \cite%
{abdullaev2011solitons,nixon2012stability,moiseyev2011crossing,lumer2013nonlinearly,jisha2014nonlocal,wang2016two}%
, time-dependent harmonic-Gaussian potential~\cite{yan2015solitons},
anharmonic sextic double well~\cite{wen2015dynamical}, and others. Various
effective potentials have been developed in photonics~\cite%
{suchkov2016nonlinear} and other physical media \cite%
{burlak2013stability,bludov2013stable,fortanier2014dipolar,dizdarevic2015cusp,dai2017localized, dai2017dynamics, li2016asymmetric,Lombard2017,Mihalache2018a,Mihalache2018b,Lombard2018}%
. Families of non-parity-time-symmetric solitons are also admitted in parity-time-symmetric potentials~\cite{yang2014can}.

Recently, a broad variety of $\mathcal{PT}$-symmetric nonlinear localized
modes and their stability were addressed in the framework of generalized
nonlinear Schr\"{o}dinger (NLS) equations with variable group-velocity
coefficients~\cite{yan2016stable,chen2018one}, third-order dispersion~\cite%
{chen2016solitonic} and position-dependent effective mass~\cite%
{chen2017families}, as well as derivative \cite{chen2017stable}, nonlocal~%
\cite{wen2017solitons}, and other generalized nonlinearities~\cite%
{yan2017nonlinear}, and in the three-wave-interaction model~\cite%
{shen2018effect}. Localized states in a nearly-$\mathcal{PT}$-symmetric
Ginzburg-Landau model were considered too~\cite{chen2018impact}. Stable
solitons of peakon and flat-top types can be maintained by $\mathcal{PT}$%
-symmetric potentials~combining terms with $\delta (x)$ and $|x|$ \cite%
{chen2019stable}. Also, stable solitons pinned to attraction centers
combined with parity-time-symmetric $\delta $-functional dipoles in the 1D
system with the critical and supercritical self-focusing nonlinearity were
reported~\cite{Mayteevarunyoo2013,wang2019attraction}. Stable solitons
in the 2D $\PT$-symmetric potentials were found in \cite{kartashov2015topological} and
examples of stable asymmetric solitons in 3D settings were provided
in \cite{kartashov2016three}. More vortex solitons in various physical contexts were reviewed in Ref.~\cite%
{malomed2019vortex}. Theoretical and experimental findings for diverse
nonlinear modes in $\mathcal{PT}$-symmetric media are summarized in several
recent reviews \cite%
{konotop2016nonlinear,suchkov2016nonlinear,he2014localized,
mihalache2017multidimensional,el-ganainy2018non,christodoulides2018parity,malomed2019nonlinear,mihalache2021localized}%
.

Nonlinear modes supported by $\mathcal{PT}$-symmetric potentials remain a
relevant topic for ongoing studies~\cite{konotop2016nonlinear}. In this paper, we develop an analytical
treatment and numerical investigation of nonlinear localized modes,
represented in an integral-convolution form, in one-, two-, and
three-dimensional (1D, 2D, and 3D) NLS equations with a novel category of $%
\mathcal{PT}$-symmetric generalized Scarf-II potentials.

The rest of this paper is organized as follows. First, the linear spectral problem with the $\mathcal{PT}$-symmetric potential
is
considered in Sec. 2. Next, Sec. 3 addresses 1D, 2D and 3D nonlinear
localized modes and their dynamical characteristics, such as stability and
excitation in $\mathcal{PT}$-symmetric GS-II double-well potentials in the
presence of the Kerr nonlinearity. Importantly, the corresponding $n$%
-dimensional solitons are obtained in an exact form. In Sec. 4, another type
of $\mathcal{PT}$-symmetric GS-II potentials are produced, and the
corresponding 3D nonlinear localized modes are investigated, including their
stability. The paper is concluded by Sec. 5.

\section*{2.\thinspace\ The higher-dimensional $\mathcal{PT}$-symmetric NLS equation}

In Kerr-nonlinear optical media with complex-valued external potentials, the
transmission of optical beams is governed by the following generalized NLS
equation~written in the dimensionless form \cite{musslimani2008optical,
shi2011bright}:
\begin{equation}
i\frac{\partial \psi}{\partial z}+[\nabla_x^{2} +V({\bm x})+iW(x) +g|\psi |^{2}]\psi =0,  \label{nls}
\end{equation}%
where the complex envelope field $\psi \equiv \psi ({\bm x},z)\in \mathbb{C}[{\bm x},z]$ is the
slowly-varying light-field amplitude, ${\bm x}=(x_{1},...,x_{n})$ are
coordinates in the $n$-dimensional space, \textit{viz}., ${\bm x}=x$ and $%
(x,y)$ in the 1D and 2D waveguides, whereas in the 3D space, the additional
coordinate is the temporal one, ${\bm x}=(x,y,t)$, $\nabla _{{\bm x}}$ is
the $n$-dimensional gradient operator, $\nabla _{{\bm x}}^{2}=\sum_{j=1}^{n}%
\partial_{x_j}^{2}$ is the Laplacian operator, $z$ is the
propagation distance, and $g$ defines the self-focusing ($g=1$) or
defocusing ($g=-1$) sign of the cubic nonlinearity. The complex potential $V(%
{\bm x})+iW({\bm x})$ is $\mathcal{PT}$-symmetric under the constraints
$V({\bm x})=V(-{\bm x}),\quad W(-{\bm x})=-W({\bm x}).$
In the optical waveguide, the real part $V({\bm x})$ of the potential is
proportional to local value of refractive index, while the imaginary part $W(%
{\bm x})$ represents the gain (amplification) or loss (absorption) in the
medium.

Under the present assumptions, Eq.~(\ref{nls}) is invariant under the
combined actions of the parity- and time-reflection, defined as $\mathcal{P}:%
{\bm x}\rightarrow -{\bm x};\,\mathcal{T}:i\rightarrow -i,z\rightarrow -z$,
respectively. The invariance implies that if $\psi (x,z)$ is an exact
solution of Eq.~(\ref{nls}), so is $\mathcal{PT}\psi (x,z)=\psi ^{\ast
}(-x,-z)$. Usually, analytically available solutions of Eq.~(\ref{nls}) are $%
\mathcal{PT}$-symmetric, i.e., they are converted into themselves by the $%
\mathcal{PT}$ transform.
Eq.~(\ref{nls}) can also be written in the variational form, $i\psi
_{z}=\delta \mathcal{H}(\psi )/(\delta \psi ^{\ast })$, with the generalized
Hamiltonian
\begin{equation}
\mathcal{H}\!=\!\!\int_{\mathbb{R}^{n}}\!\!\left\{ |\nabla _{\bm x}\psi
|^{2}\!-\![V({\bm x})\!+\!iW({\bm x})]|\psi |^{2}\!-\!\frac{g}{2}|\psi
|^{4}\!\right\} \!d{\bm x},
\end{equation}%
where the asterisk stands for the complex conjugate. Because the Hamiltonian
is not real, Eq. (\ref{nls}) does not conserve the total power,
$P(z)=\int_{\mathbb{R}^{n}}|\psi ({\bm x},z)|^{2}d{\bm x}.$
The evolution equation for the power follows from Eq. (\ref{nls}):
\begin{equation}
\frac{dP(z)}{dz}=-2\int_{\mathbb{R}^{n}}W({\bm x})|\psi ({\bm x},z)|^{2}d{\bm x}%
.  \label{dP/dz}
\end{equation}%
In particular, if the local intensity, $|\psi ({\bm x},z)|^{2}$, is an even function of ${%
\bm x}$, as is usually the case for stationary modes, while the $\mathcal{PT}
$ symmetry demands $W({\bm x})$ to be an odd function of ${\bm x}$,
Eq.~(\ref{dP/dz}) yields $dP/dz=0$, that is, the total power
is conserved for such solutions.

\subsection*{2.1 \thinspace\ The analysis of nonlinear modes}

In general, Eq.~(\ref{nls}) with complex potential $V({\bm x})+iW({\bm x})$
is a non-integrable equation, as the Lax pair, which is necessary for the
integrability, is available only in rare special cases~\cite{yan2013complex}%
. In the absence of the integrability, classical methods for constructing
exact solutions, such as the bilinear Hirota technique and Darboux
transformation, cannot be used. Therefore, we here aim to find nonlinear
modes as steady-state solutions, $\psi ({\bm x},z)=\phi ({\bm x})e^{i\mu z}$
with a real propagation constant $\mu $. The wave function $\phi ({\bm x})$,
which is assumed to vanish at $|x|\rightarrow \infty $, as we look for\
localized states, 
satisfies the stationary version of Eq. (\ref{nls}):
\begin{equation}
\left( \nabla _{{\bm x}}^{2}-\mu \right) \phi =-\left[ V({\bm x})+iW({\bm x}%
)+g|\phi |^{2}\right] \phi .  \label{stat}
\end{equation}%
This complex-valued wave function $\phi(x)$ can be further expressed in the Madelung form
\begin{equation}
\phi({\bm x})=u({\bm x})e^{i\varphi({\bm x})},  \label{Madelung}
\end{equation}%
with real $u({\bm x})$ and $\varphi({\bm x})$ being, respectively, real amplitude and phase. The substitution of the Madelung ansatz into Eq. (\ref%
{stat}) leads to coupled nonlinear equations
\begin{equation} \label{eq-re}
\begin{array}{l}
 \nabla _{{\bm x}}\cdot\Big(u^2({\bm x}) \nabla_{{\bm x}}\varphi({\bm x})\Big)=-W({\bm x})u^2({\bm x}),%
\vspace{5pt} \\
\Delta _{\bm x}u({\bm x})=[-V({\bm x})-gu^{2}({\bm x})+\nabla_{{\bm x}}\varphi({\bm x})\cdot \nabla_{{\bm x}}\varphi({\bm x})+\mu ]u(%
{\bm x}).%
\end{array}%
\end{equation}%

In particular, for the 1D case: ${\bm x}\to x$, let $\varphi(x)=\int_{0}^xv(s)ds$, that is,
\begin{equation}
\phi(x)=u(x)\exp \left[i\int_{0}^xv(s)ds\right],  \no
\end{equation}%
with real $u(x)$ and $v(x)$ being, respectively, real amplitude and
hydrodynamic velocity. Then system (\ref{eq-re}) becomes
\begin{equation} \no
\begin{array}{l}
v(x)=-u^{-2}(x)\displaystyle\int_{0}^xW(s)u^{2}(s)ds,
\vspace{5pt} \\
\partial_x^2u(x)=[-V(x)-gu^{2}(x)+v^{2}(x)+\mu ]u(x).%
\end{array}%
\end{equation}%


\subsection*{2.2 \thinspace\ Representation of general solutions}

To seek soliton solutions of Eq.~(\ref{stat}), we restrict them to
the \textit{Schwartz space} of rapidly-decreasing functions~\cite%
{cazenave2003semilinear}
\bee
\phi ({\bm x})\in \wp (\mathbb{R}^{n})=\{f({\bm x})\in C^{\infty }(\mathbb{%
R}^{n})|\lim_{|{\bm x}|\rightarrow \infty }{\bm x}^{\alpha }\partial ^{p}f({%
\bm x})=0, \, \forall \alpha ,p\in \mathbb{N}^{n}\}, \no
\ene
where ${\bm x}^{\alpha }\partial ^{p}f({\bm x})\equiv {x_{1}}^{\alpha
_{1}}\cdots {x_{n}}^{\alpha _{n}}\partial _{x_{1}}^{p_{1}}\cdots \partial
_{x_{n}}^{p_{n}}f({\bm x})$ and $C^{\infty }(\mathbb{R}^{n})$ represents the
space consisting of infinitely continuous differentiable complex-valued
functions.

We assume that the trapping potential selects solutions with $\mu >0$. The
Green-function equation corresponding to Eq.~(\ref{stat}) is
\begin{equation}
\nabla _{{\bm x}}^{2}G-\mu G=\delta ({\bm x})=\delta (x_{1})\delta
(x_{2})\cdots \delta (x_{n}).  \label{gf}
\end{equation}%
Applying the Fourier transform to Eq.~(\ref{gf}) and then the inverse
transform leads to the expression for the Green's function (fundamental
solution),
\begin{equation}
G({\bm x})=-\frac{1}{(2\pi )^{n}}\int_{\mathbb{R}^{n}}\frac{e^{i{\bm x}\cdot %
\bm{\xi}}}{|\bm{\xi}|^{2}+\mu }d\bm{\xi}.  \label{Green}
\end{equation}%
Then, solutions of Eq.~(\ref{stat}) can be represented in the form of the
convolution with the Green's function:
\begin{equation}
\phi _{G}({\bm x})=-G({\bm x})\ast \left[ V({\bm x})\phi +iW({\bm x})\phi
+g|\phi |^{2}\phi \right] .  \label{fs}
\end{equation}%
%
%
%

Further, Eq.~(\ref{fs}) can be written in a more specific form:
\begin{align} \label{gs}
\tilde{\phi}\!=\!C_{1}\exp {\left( \sum_{k=1}^{n}\beta _{k}x_{k}\right) }%
+C_{2}\exp {\left( \sum_{k=1}^{n-1}\beta _{k}x_{k}\!-\!\beta _{n}x_{n}\right) }\!-\!\int_{\mathbb{R}^{n}}\!\!G({\bm x}-{\bm y})\!\left[ V({\bm y})\!+\!iW({\bm y})\!+\!g|\phi ({\bm y})|^{2}\right] \phi ({\bm y})d{\bm y}.
\end{align}%
%
%
%
%
%
%
where $\beta _{n}=\sqrt{\mu -\sum_{k=1}^{n-1}\beta _{k}^{2}}$, and $%
C_{1},C_{2},\beta _{k}(k=1,2,\cdots ,n-1)$ are arbitrary complex constants.
In particular,

\begin{itemize}
\item {} In the 1D setting, Eq. (\ref{gs}) reduces to
\begin{align}  \label{gs1}
& \tilde{\phi}(x)=C_{1}e^{\sqrt{\mu }x}+C_{2}e^{-\sqrt{\mu }x}-\int_{\mathbb{%
R}^{n}}\!G(x-x^{\prime })  \left[ V(x^{\prime })\!+\!iW(x^{\prime })\!+\!g|\phi(x^{\prime})|^2\right] \phi (x^{\prime })dx^{\prime }.
\end{align}

\item {} In the 2D case, Eq. (\ref{gs}) becomes
\begin{align}\label{gs2}
& \tilde{\phi}(x,y)=C_{1}e^{\beta x+\sqrt{\mu -\beta ^{2}}y}+C_{2}e^{\beta x-%
\sqrt{\mu -\beta ^{2}}y} -\int_{\mathbb{R}^{2}}\!\!G(x-x^{\prime },y-y^{\prime })\!\left[
V(x^{\prime },y^{\prime })\!+\!iW(x^{\prime },y^{\prime })\right.  \notag \\
& \qquad \left. +g|\phi (x^{\prime },y^{\prime})|^2\right] \phi (x^{\prime},y^{\prime})dx^{\prime }dy^{\prime },
\end{align}%
where $\beta $ is an arbitrary real constant.
\end{itemize}

\subsection*{2.3 \thinspace\ The iteration scheme for constructing numerical
nonlinear modes}

\ Define the Fourier transform $\mathcal{F}$ of $\phi ({\bm x})$ by
\begin{equation} \no
\hat{\phi}(\bm{\xi})=\mathcal{F}[\phi ({\bm x})]=\frac{1}{(\sqrt{2\pi })^{n}}%
\int_{\mathbb{R}^{n}}\phi ({\bm x})e^{-i{\bm x}\cdot \bm{\xi}}d\bm{\xi}.
\end{equation}%
Applying this transform to Eq.~(\ref{stat}) yields
\begin{equation}
\hat{\phi}(\bm{\xi})=\frac{\mathcal{F}[V({\bm x})\phi ]+i\mathcal{F}[W({\bm x%
})\phi ]+g\mathcal{F}[|\phi |^{2}\phi ]}{|\bm{\xi}|^{2}+\mu }.  \label{phi-f}
\end{equation}%
To secure the convergence of the iterative scheme, a new field variable is
usually introduced, $\phi ({\bm x})=\gamma \Phi ({\bm x})$, where $\gamma $
is a nonzero constant to be determined, hence $\hat{\phi}(\bm{\xi})=\gamma
\hat{\Phi}(\bm{\xi})$. Then Eq.~(\ref{phi-f}) becomes
\begin{equation}
\hat{\Phi}(\bm{\xi})=\frac{\mathcal{F}[V\Phi ]+i\mathcal{F}[W\Phi ]+g%
\mathcal{F}[|\gamma |^{2}|\Phi |^{2}\Phi ]}{|\bm{\xi}|^{2}+\mu }\triangleq
G_{\gamma }(\bm{\xi}). \no
\end{equation}%
Multiplying it by $\hat{\Phi}^{*}(\bm{\xi})$ and integrating over the
entire $\bm{\xi}$ space leads to an equation for $\gamma $,
\begin{equation}
\int_{\mathbb{R}^{n}}\left( |\hat{\Phi}(\bm{\xi})|^{2}-\hat{\Phi}^{\ast }(%
\bm{\xi})G_{\gamma }(\bm{\xi})\right) d\bm{\xi}=0.  \label{gamma}
\end{equation}%
Thus, the iteration scheme of for constructing numerical nonlinear modes is
defined as
\begin{equation}  \label{ite}
\hat{\Phi}_{m+1}\!=\!\frac{\mathcal{F}[V\Phi _{m}]\!+\!i\mathcal{F}[W\Phi
_{m}]\!+\!g\mathcal{F}[|\gamma _{m}|^{2}|\Phi _{m}|^{2}\Phi _{m}]}{|\bm{\xi}%
|^{2}+\mu },
\end{equation}%
where $\gamma _{m}$ is determined by Eq.~(\ref{gamma}) with $\Phi =\Phi
_{m},\,\gamma =\gamma _{m}$. 

Other numerical algorithms can also be employed to solve Eq.~(\ref{stat})
with the zero boundary conditions, such as the classical shooting method for
1D fundamental and higher-order solitons, spectral renormalization method
for multidimensional self-localized states~\cite{ablowitz2005spectral}, the
modified squared-operator iteration method~\cite{yang2007universally}, as
well as the Newton's conjugate-gradient method for multidimensional and
higher-order nonlinear modes~\cite{yang2009newton}. We here make use of the modified squared-operator iteration method to search for higher-order numerical nonlinear modes, as
it allows us to reach higher precision, and is convenient for the application in higher dimensions.
Besides, the combination of the
Fourier pseudospectral method and split-step algorithm are used to simulate the propagation of
nonlinear modes~\cite{yang2010nonlinear}.

While most results in this area are obtained in the fully numerical form,
for particular types of $\mathcal{PT}$ potential wells, exact nonlinear modes
play a crucial role in seeking  families of stable numerical solitons
(see, e.g., Refs.~\cite{yan2015spatial,wen2015dynamical,yan2015solitons}).

\subsection*{2.4 \thinspace\ The stability analysis of nonlinear modes}

When nonlinear localized modes of Eq.~(\ref{nls}) are obtained in the form
of expression $\psi ({\bm x},z)=\phi ({\bm x})e^{i\mu z}$, exact or numerical, one can analyze their
linear stability by perturbing them as follows:
\begin{equation}
\psi ({\bm x},z)=\left[ \phi ({\bm x})+\varepsilon \left( f_{1}({\bm x}%
)e^{i\nu z}+f_{2}^{\ast }({\bm x})e^{-i\nu ^{\ast }z}\right) \right] e^{i\mu
z},\,\,\,
\end{equation}%
where $|\varepsilon |\ll 1$ is an amplitude of the small perturbation, both $%
f_{1}({\bm x})$ and $f_{2}({\bm x})$ are perturbation eigenfunctions, and $%
\nu $ is the (in)stability eigenvalue.
Substituting the perturbed solutions in Eq.~(\ref{nls}) and linearizing it
with respect of $\varepsilon $, gives rise to the following linear-stability
eigenvalue problem
\begin{equation}
\left(
\begin{array}{cc}
\tilde{L}-\nu & g\phi ^{2}({\bm x}) \vspace{0.05in} \\
-g\phi^{*2}({\bm x}) & -\tilde{L}^{\ast }-\nu%
\end{array}%
\right) \left(
\begin{array}{c}
f_{1}({\bm x})\vspace{0.05in} \\
f_{2}({\bm x})%
\end{array}%
\right) =0  \label{sta}
\end{equation}%
with 
$\tilde{L}=\nabla _{{\bm x}}^{2}+[V({\bm x})+iW({\bm x})]+2g|\phi ({\bm x}%
)|^{2}-\mu $. 

The steady-state nonlinear modes $\phi ({\bm x})e^{i\mu z}$ are linearly
unstable once $\nu $ has a negative imaginary part, otherwise they are
stable. The celebrated Fourier collocation method is a practical and
efficient tool to compute the entire linear-stability spectrum \cite%
{yang2010nonlinear}. While the linear-stability analysis roughly predicts a parameter range
corresponding to stability of nonlinear modes, it provides only necessary
but not sufficient condition for the true nonlinear stability. The full
stability should be then tested by means of direct numerical simulations of
the perturbed evolution.

\subsection*{2.5 \, $\mathcal{PT}$-symmetry breaking in the GS-II potential}

In this subsection, we concentrate on the $\mathcal{PT}$-symmetric breaking
phenomena in the linear regime, based on the equation
\begin{equation}
\mathcal{L}\Psi (x)=\lambda \Psi (x),\quad \mathcal{L}=-\frac{d^{2}}{dx^{2}}%
\!-\!V(x)\!-\!iW(x),  \label{lp}
\end{equation}%
with the 1D $\mathcal{PT}$-symmetric GS-II potential taken as
\begin{equation}
\begin{array}{l}
V(x)=v_{1}\,\mathrm{sech}^{2}x+v_{2}\,\mathrm{sech}^{2\alpha }x,\quad
W(x)=w_{0}\,\mathrm{sech}^{\alpha }x\tanh x,%
\end{array}
\label{GS}
\end{equation}%
where $\alpha >0$ (the \textit{localization parameter}),\thinspace\ $v_{1}>0$%
, $v_{2}$, and $w_{0}$ are real parameters, while $\lambda $ and $\Psi (x)$
represent the discrete spectrum and the corresponding bound-state
eigenfunctions. This $\mathcal{PT}$-symmetric potential with $\alpha =1$ and
$v_{1}>0,\,v_{2}=0$ (or $v_{1}>-v_{2}\not=0$) reduces to the classical
Scarf-II potential~\cite{gendenshtein1983derivation,musslimani2008optical}.
Another version of the GS-II potential, quite different from the one written
in Eq. (\ref{GS}), was considered in Ref. \cite{chen2017families}.

To find exact nonlinear localized modes produced by Eq. (\ref{stat}), in
what follows we focus on a special case of potential (\ref{GS}) with
\begin{equation}
v_{1}=\alpha (\alpha +1),\quad v_{2}=w_{0}^{2}/(9\alpha ^{2})-g\phi _{0}^{2},
\label{AB}
\end{equation}%
where $\phi _{0}$ is a new real-valued parameter in place of $v_{2}$. In
this case, coefficient $w_{0}$ independently determines the strength of the
gain-loss distribution $W(x)$. Once both $\alpha $ and $w_{0}$ are fixed,
one can choose any value of $v_{2}$ in Eq.~(\ref{AB}) by means of adjusting
coefficient $\phi _{0}$. Note that when the gain-loss strength $w_{0}$ or
potential index $\alpha $ changes, both real and imaginary parts of the
GS-II potential (\ref{GS}) alter, which may lead to several distinct $%
\mathcal{PT}$-symmetry-breaking scenarios.

\begin{figure}[t]
\begin{center}
\vspace{0.05in} \hspace{-0.05in}{\scalebox{0.56}[0.56]{%
\includegraphics{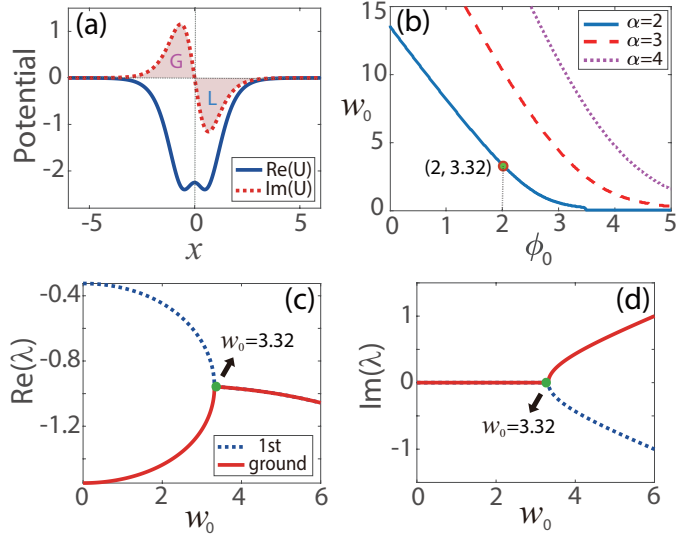}}}
\end{center}
\par
\vspace{-0.2in}
\caption{{\protect\small (color online). (a) The profile of 1D $\mathcal{PT}$%
-symmetric GS-II potential $U=-V(x)-iW(x)$, defined by Eqs. (\protect\ref{GS}%
) and (\protect\ref{AB}) with $\protect\alpha =2,\protect\phi _{0}=2,w_{0}=3$%
, and }$g=1${\protect\small ,} {\protect\small where `G' (`L') denotes the
gain (loss) region. (b) Boundaries of the $\mathcal{PT}$-symmetry breaking
in the $(\protect\phi _{0},w_{0})$ plane of the linear system (\protect\ref%
{lp}) [with no cubic term in Eq. (\protect\ref{nls})] for $\protect\alpha %
=2,3,4$. The }$\mathcal{PT}${\protect\small \ symmetry is unbroken under the
boundaries. (c) Real and (d) imaginary parts of eigenvalues $\protect\lambda
$ of the two lowest states as a function of $w_{0}$, for $\protect\alpha =2,%
\protect\phi _{0}=2$. }}
\label{spe1}
\end{figure}

Without loss of generality, we assume $\phi _{0}\geq 0,w_{0}\geq 0$, and $%
g=1 $ (the latter condition implies that the Kerr nonlinearity is
self-focusing), unless specified otherwise. For illustration, a
representative $\mathcal{PT}$-symmetric double-well GS-II potential, as well
as the corresponding gain and loss regions are exhibited in Fig.~\ref{spe1}%
(a).

Next, we look for all-real spectra in $\mathcal{PT}$-symmetric GS-II
potential (\ref{GS}) by solving the linear eigenvalue problem (\ref{lp}).
Note that, with $\alpha =1$, the GS-II potential (\ref{GS}) with
coefficients (\ref{AB}) reduces to the same form as considered in Ref. \cite%
{chen2020soliton}, where the exact condition of the persistence of the
unbroken $\mathcal{PT}$-symmetric phases was derived:
\begin{equation}
\phi _{0}\leq \frac{|2w_{0}-9|}{6}.
\end{equation}%
At $\alpha \neq 1$, we have found the $\mathcal{PT}$-symmetry-breaking
boundary by means of the numerical Fourier spectral method. The boundaries
are displayed for $\alpha =2,3,4$ in the first quadrant of the $(\phi
_{0},w_{0})$ parameter plane in Fig.~\ref{spe1}(b). It is seen that the
boundaries have a monotonous shape, approaching to zero with the increase of $%
\phi _{0}$. This happens because, as $\phi _{0}$ increases, hence $v_{2}$
decreases in Eq.~(\ref{AB}), then the real part $-V_{0}(x)$ of the GS-II
potential (\ref{GS}) gradually turns from a potential well into a barrier,
thereby leading to the emergence of complex eigenvalues and dramatic
shrinkage of existence region for fully-real spectra. On the other hand, it
is also seen that, with the growth of $\alpha $, the $\mathcal{PT}$%
-symmetry-breaking boundary rises higher, helping to expand the region of
the unbroken $\mathcal{PT}$ symmetry. In fact, due to the growth of $\alpha $%
, the potential well $V(x)$ gets deeper, while the gain-loss distribution $%
W(x)$ becomes tighter, hence more likely resulting in real eigenvalues.
Thus, a conclusion is that the range of the totally-real spectra that the
GS-II potential (\ref{GS}) produces can be enlarged by increasing $\alpha $.


It follows from Fig.~\ref{spe1}(b) that, for the fixed $\phi _{0}=2$, the $%
\mathcal{PT}$ symmetry-breaking threshold point is $w_{0}=3.32$. To
illustrate the picture in detail, the evolution of the eigenvalues of the
two lowest eigenstates following the change of coefficient $w_{0}$ in Eq. (%
\ref{AB}) is shown in Figs. \ref{spe1}(c,d), where the eigenvalues remain
completely real at $w<w_{0}=3.32$, while at $w>w_{0}$ they form a
complex-conjugate pair, produced by the bifurcation that takes place at $%
w=w_{0}$. In fact, the formal solutions with complex eigenvalues represent
the so-called ghost states (that do not exist as stationary ones, but may
represent long-lived unsteady ones~\cite%
{cartarius2012nonlinear,rodrigues2012pt,susanto2018snakes}).

Similar $\mathcal{PT}$-symmetry-breaking phenomenology occurs in the
multi-dimensional linear system. However, the key problem is how to
generalize the GS-II potential (\ref{GS}) to the $n$-dimensional case, so as
to be able to find exact nonlinear localized modes. As an example, we give
the corresponding 2D GS-II potential:
\begin{equation}  \label{GS-2D}
\begin{array}{l}
V(x,y)\!=\!\mathop{\sum}\limits_{x_{j}}(v_1\mathrm{sech}%
^{2}x_{j}\!+\!v_{21}\mathrm{sech}^{2\alpha }\!x_{j})\!+\!v_{22}\!\mathop{\prod}\limits_{x_{j}}\mathrm{sech}^{2\alpha }\!x_{j},%
\vspace{5pt} \\
W(x,y)=w_{0}\mathop{\sum}\limits_{x_{j}}\mathrm{sech}^{\alpha
}\!x_{j}\tanh x_{j},%
\end{array}%
\end{equation}%
where $x_j=x,y$, and $v_1,\, v_{21}$ and $v_{22}$ are real constants with the 2D
counterpart of Eq.~(\ref{AB}) being
\begin{equation}
v_1=\alpha (\alpha +1),\quad v_{21}=w_{0}^{2}/(9\alpha ^{2}),\quad
v_{22}=-g\phi _{0}^{2}.  \label{AB12}
\end{equation}%
%
%
%
$\mathcal{PT}$-symmetry-breaking boundaries in this 2D system can be
identified in the numerical form, similar to how it is shown above for 1D.

Lastly, the 3D GS-II potential can be obtained too, in the following form:
\begin{equation}
\begin{array}{l}
V(x,y,t)\!=\!\mathop{\sum}\limits_{x_{j}}(v_1\mathrm{sech}%
^{2}x_{j}\!+\!v_{21}\mathrm{sech}^{2\alpha }\!x_{j})\!+\!v_{22}\!\mathop{\prod}\limits_{x_{j}}\mathrm{sech}^{2\alpha }\!x_{j},%
\vspace{5pt} \\
W(x,y,t)=w_{0}\mathop{\sum}\limits_{x_{j}}\mathrm{sech}^{\alpha
}\!x_{j}\tanh x_{j},%
\end{array}
\label{GS-3D}
\end{equation}%
where $x_j=x,y,t$, and $v_{1},\,v_{21},\,v_{22},\,w_{0}$ are determined by Eq.~(\ref{AB12}).

\section*{3 \thinspace\ Nonlinear modes and stability in the
$\mathcal{PT}$-symmetric GS-II potential}

In this section, we firstly produce particular solutions for 1D, 2D and 3D
exact nonlinear localized modes in the nonlinear system (\ref{nls}) with the
GS-II potential (\ref{GS}). Then we numerically explore families of
fundamental solitons including such exact modes, and the corresponding
high-order nonlinear modes. The dynamical stability of these states is
investigated in detail. In the end, we summarize the results obtained for
the nonlinear localized states in the $n$-dimensional $\mathcal{PT}$%
-symmetric GS-II potential.

\subsection*{3.1 \thinspace\ 1D solitons and stability}

\subsubsection*{3.1.1 \thinspace\ Analytical solitons and the linear
stability}

Particular solutions of one-dimensional equation (\ref{stat}) with the Kerr
nonlinearity and $\mathcal{PT}$-symmetric GS-II potential (\ref{GS}),
subject to constraint Eq.~(\ref{AB}), can be found, for the propagation
constant $\mu =\alpha ^{2}$, as
\begin{equation}
\phi (x)=\phi _{0}\,\mathrm{sech}^{\alpha }\!x\exp \!\left( \dfrac{iw_{0}}{%
3\alpha }\int_{0}^{x}\mathrm{sech}^{\alpha }sds\right) .  \label{nlm1}
\end{equation}%
It is worthy to note that $\phi (x)\in \wp (\mathbb{R})$ and local intensity
$|\phi (x)|^{2}=\phi _{0}^{2}\,\mathrm{sech}^{2\alpha }\!x$ of these exact
solutions depends only on $\phi _{0}$ and $\alpha $, while the gain-loss
strength $w_{0}$ does not appear in the solution.


\begin{figure}[t]
\begin{center}
\vspace{0.05in} \hspace{-0.05in}{\scalebox{0.5}[0.5]{%
\includegraphics{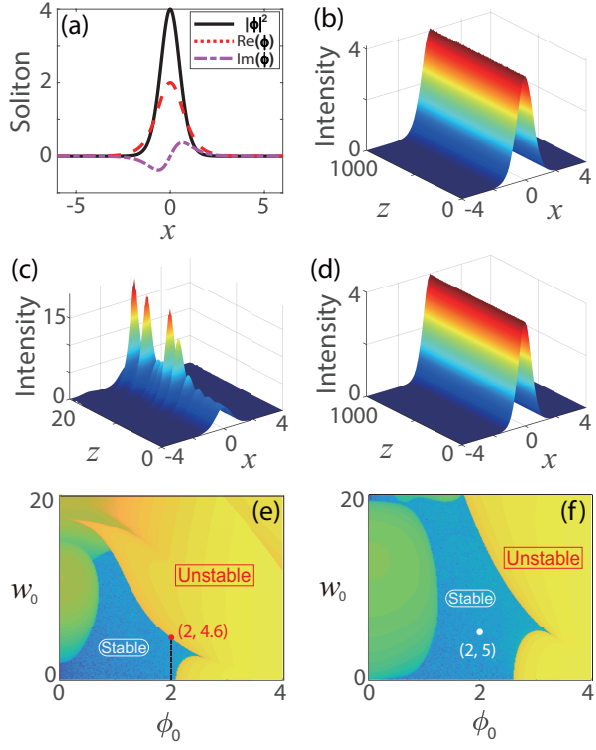}}}
\end{center}
\par
\vspace{-0.2in}
\caption{{\protect\small (a) The 1D exact fundamental soliton with $\protect%
\alpha =2,\protect\phi _{0}=2,w_{0}=3$, and (b) its stable evolution. (c)
Unstable propagation: $\protect\alpha =2,\protect\phi _{0}=2,w_{0}=5$; (d)
stable propagation: $\protect\alpha =3,\protect\phi _{0}=2,w_{0}=5$. The
linear-stability map of exact solitons (\protect\ref{nlm1}) in the $(\protect%
\phi _{0},w_{0})$ parameter plane: $\protect\alpha =2$ (e), and $\protect%
\alpha =3$ (f), where the blue (or dark, in the black-and-white rendition)
regions host stable solitons, while other regions, such as yellow and green ones (light, in the
black-and-white format), are populated by unstable solitons. The
nonlinearity coefficient is $g=1$ here.}}
\label{exa1}
\end{figure}

Next, we focus on the analysis of the exact nonlinear localized mode (\ref%
{nlm1}) in two cases, as follows.

\textit{Case 1.}\thinspace\ At $\alpha =1$, the $\mathcal{PT}$-symmetric
GS-II potential (\ref{GS}) turns into the remarkable Scarf-II potential,
\begin{equation}
V(x)=v_{0}\,\mathrm{sech}^{2}x,\quad W(x)=w_{0}\,\mathrm{sech}x\tanh x,
\end{equation}%
with $v_{0}=v_{1}+v_{2}=w_{0}^{2}/9-g\phi _{0}^{2}+2.$ Then, soliton (\ref%
{nlm1}) of Eq.~(\ref{stat}) with $\mu =1$ becomes~\cite{chen2020soliton}
\begin{equation}
\phi (x)=\sqrt{A/g}\,\mathrm{sech}x\,e^{iw_{0}\tan ^{-1}(\sinh x)/3},\quad
A=\left( w_{0}^{2}/9-v_{0}+2\right),\quad gA >0.
\end{equation}%

\begin{itemize}
\item For the focusing Kerr nonlinearity ($g=1$), the soliton (\ref{nlm1})
is $\phi (x)=\sqrt{A}\,\left( \mathrm{sech}x\right) \exp %
\left[ \left( iw_{0}/3\right) \tan ^{-1}(\sinh x)\right] $ \cite%
{musslimani2008optical}. 

\item For the defocusing case ($g=-1$), the soliton (\ref{nlm1})
becomes $\phi (x)=\sqrt{-A}\mathrm{sech}\exp \left[ \left( iw_{0}/3\right) \tan ^{-1}(\sinh x)\right] $
\cite{shi2011bright}.
\end{itemize}

\textit{Case 2.}\thinspace\ At $\alpha >0$ with $\alpha \neq 1$, the
proposed $\mathcal{PT}$-symmetric GS-II potential (\ref{GS}) essentially
extends the original Scarf-II potential, and the width of the corresponding
bright soliton (\ref{nlm1}) changes. As $0<\alpha <1$, the width becomes
larger, while as $\alpha >1$ it is smaller. 

To facilitate the consideration, we start our analysis for $\alpha =2$ as the
reference value. Under the GS-II potential (\ref{GS}) in Fig.~\ref{spe1}(a),
a representative exact ground-state soliton with one hump is produced by
Eq.~(\ref{nlm1}) (see Fig.~\ref{exa1}(a)). We have found that, for the
strength of the gain-loss distribution $w_{0}=3$, the soliton is stable in
long direct simulations with $2\%$ white noise added to the soliton, see
Fig.~\ref{exa1}(b). If this parameter increases to $w_{0}=5$, exceeding a
certain threshold value, the exact nonlinear mode (\ref{nlm1}) becomes
unstable, quickly spreading out in the simulations, see Fig.~\ref{exa1}(c).
This result is naturally explained by the fact that too large gain-loss
force is adverse to the stability of the soliton. However, in this case, the
exact soliton (\ref{nlm1}) regains its stability again for $\alpha =3$, see
Fig.~\ref{exa1}(d). In fact, the growth of $\alpha $ makes the potential
well deeper, thus helping to stabilize the soliton.

To further confirm these results, the linear-stability analysis can be
utilized to seek for a region where the solitons is stable is located in the
$(\phi _{0},w_{0})$ parameter plane. Accordingly, the linear stability is
determined by the largest absolute value of imaginary parts of the
eigenvalue $\nu $ in Eqs. (\ref{nlm1}) and (\ref{sta}), and the logarithmic
scale is used, displaying the results in the form of $\lg [\max |\Im (\nu
)|] $. At $\alpha =2$ and $3$, we thus find a broad blue area of stable
solitons, which is shown in Figs.~\ref{exa1}(e,f). Note that the
soliton-stability area expands as $\alpha $ increases. For example, point $%
(\phi _{0},w_{0})=(2,5)$ falls in the unstable region for $\alpha =2$,
belonging to the stability region for $\alpha =3$, which explains the
propagation picture plotted in Figs.~\ref{exa1}(c,d). Further, one can
observe in Fig.~\ref{exa1}(e) that, if $\phi _{0}=2$ is fixed, the
linear-stability threshold is $w_{0}\approx 4.6$, below which all the exact
solitons are stable. This conclusion agrees well with the findings presented
in Figs.~\ref{exa1}(b,c).

By comparing Figs.~\ref{exa1}(e) and~\ref{spe1}(b), we conclude that the
presence of the self-focusing Kerr nonlinearity can further extend the range
of fully real spectra in the $\mathcal{PT}$-symmetric GS-II potential (\ref%
{GS}). In other words, the stability range of the localized modes can be
broadened with the help of the appropriate nonlinearity. For instance, in
Fig.~\ref{exa1}(e) we observe a stable soliton under the action of the Kerr
nonlinearity at $(\phi _{0},w_{0})=(2,4)$, while the $\mathcal{PT}$ symmetry
is broken in the corresponding linear system, see Fig.~\ref{spe1}(b).

Furthermore, the power flow (Poynting vector) associated with the nonlinear
localized mode (\ref{nlm1}) is
\begin{equation}
S(x)=\frac{i}{2}(\phi \phi _{x}^{\ast }-\phi ^{\ast }\phi _{x})=\frac{\phi
_{0}^{2}w_{0}}{3\alpha }\mathrm{sech}^{3\alpha }x.
\end{equation}%
It can be seen that its positive or negative sign is solely determined by
the sign of the gain-loss strength $w_{0}$. Thus, at $w_{0}>0$, the power
always flows from the gain to loss region, as usual. In addition, at $\alpha
=2$, the total power of the exact nonlinear mode (\ref{nlm1}) is
\begin{equation}
P=\int_{-\infty }^{+\infty }|\psi (x,z)|^{2}dx=\frac{4}{3}\phi _{0}^{2}.
\end{equation}

\begin{figure}[t]
\begin{center}
\vspace{0.05in} \hspace{-0.05in}{\scalebox{0.6}[0.5]{%
\includegraphics{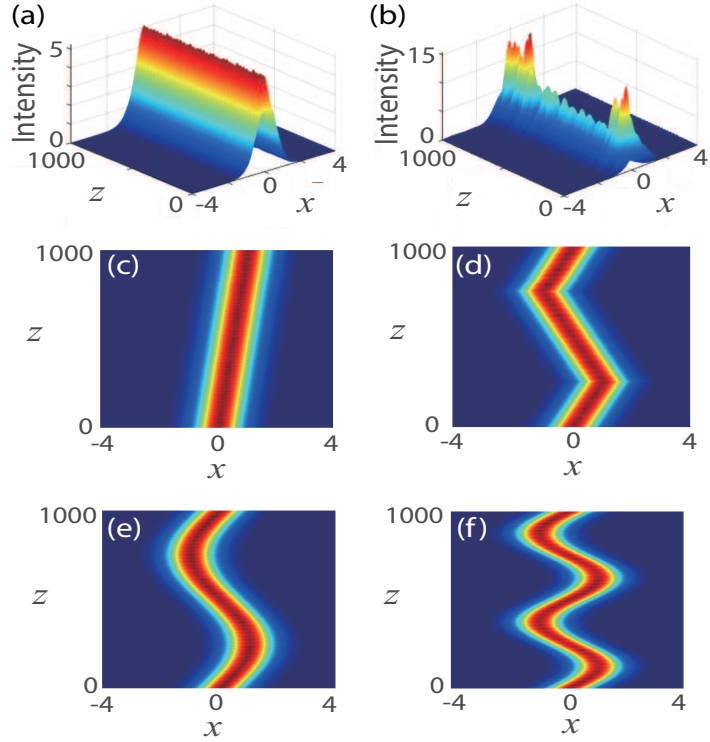}}}
\end{center}
\par
\vspace{-0.25in}
\caption{{\protect\small The propagation of nonlinear mode (\protect\ref%
{nlm1}) initially shifted by $x_{0}=0.1$ (a), or $x_{0}=0.5$ (b). Perturbed
propagation patterns of nonlinear mode (\protect\ref{nlm1}) for different
GS-II potentials (\protect\ref{nlm1}): (c) $\tilde{x}_{0}(z)=z/z_{\max }$,
(d) $\tilde{x}_{0}(z)$ determined by Eq.~(\protect\ref{plf}), (e) $\tilde{x}%
_{0}(z)=\sin (2\protect\pi z/z_{\max })$, (f) $\tilde{x}_{0}(z)=\sin (4%
\protect\pi z/z_{\max })$. Other parameters are $g=1,\protect\alpha =2,%
\protect\phi _{0}=2,w_{0}=3$. }}
\label{ps}
\end{figure}

\subsubsection*{3.1.2 \thinspace\ Propagation patterns of solitons under
perturbations}

Another natural question concerning the dynamics of the solitons is their
motion initiated by a shift, $x\rightarrow x-x_{0}$, of the exact soliton (%
\ref{nlm1}) from the equilibrium position, i.e., taking the initial
condition as
\begin{equation}
\breve{\phi}=\phi _{0}\,\mathrm{sech}^{\alpha }(x-x_{0})\,\exp {\!\left(
\frac{iw_{0}}{3\alpha }\int_{0}^{x-x_{0}}\!\!\mathrm{sech}^{\alpha
}sds\!\right) }.  \label{nlm1p}
\end{equation}%
As a result, we have found that only for sufficiently small $x_{0}$, such as
$x_{0}=0.1$, the soliton remain stable in the course of long propagation, as
shown in Fig.~\ref{ps}(a). For larger $x_{0}$, such as $x_{0}=0.5$, the
initially shifted soliton becomes unstable, see Fig.~\ref{ps}(b).

A related question is a possibility to control the transmission of the beam
by making the transverse position of the potential well be a function of the
propagation distance. To this end, we modify the GS-II potential given by
Eqs.~(\ref{GS}) and (\ref{AB}) with $x\to x-\tilde{x}_{0}(z)$. 
In this context, we consider linear and periodic functions for $\tilde{x}%
_{0}(z)$.

\begin{itemize}
\item The linear shift: for $\tilde{x}_{0}(z)=z/z_{\max }$, where $z_{\max }$
is the maximum of the propagation distance, the accordingly perturbed
soliton (\ref{nlm1}) propagates stably along a straight line from the
original position, as shown in Fig.~\ref{ps}(c). If $\tilde{x}_{0}(z)$ is
chosen as a piecewise-linear function,
\begin{equation}
\tilde{x}_{0}(z)=%
\begin{cases}
\dfrac{4z}{z_{\max }}, & \!\!\text{$0\leq 4z\leq z_{\max }$},\vspace{0.05in}
\\
2-\dfrac{4z}{z_{\max }}, & \!\!\text{$z_{\max }<4z\leq 3z_{\max }$},\vspace{%
0.05in} \\
\dfrac{4z}{z_{\max }}-4, & \!\!\text{$3z_{\max }<4z\leq 4z_{\max }$},%
\end{cases}
\label{plf}
\end{equation}%
the soliton propagates stably in the potential given by Eqs.~(\ref{GS}) and (%
\ref{AB}) with $x\rightarrow x-\tilde{x}_{0}(z)$, regardless of the
unsmoothness of the modulation pattern (\ref{plf}), see Fig.~\ref{ps}(d).

\item Under the action of the periodic shift, $\tilde{x}_{0}(z)=\sin (2\pi
z/z_{\max })$, or one with the double frequency, $\tilde{x}_{0}(z)=\sin
(4\pi z/z_{\max })$, the accordingly perturbed soliton (\ref{nlm1})
maintains stable snaking transmission, see Figs.~\ref{ps}(e) and \ref{ps}(f).
\end{itemize}

Thus, it is possible to conclude that the soliton keeps its stability in the
spatially modulated potential, with the soliton's center staying close to
the midpoint of the moving potential well.

\begin{figure}[t]
\begin{center}
\vspace{0.05in} \hspace{-0.05in}{\scalebox{0.55}[0.55]{%
\includegraphics{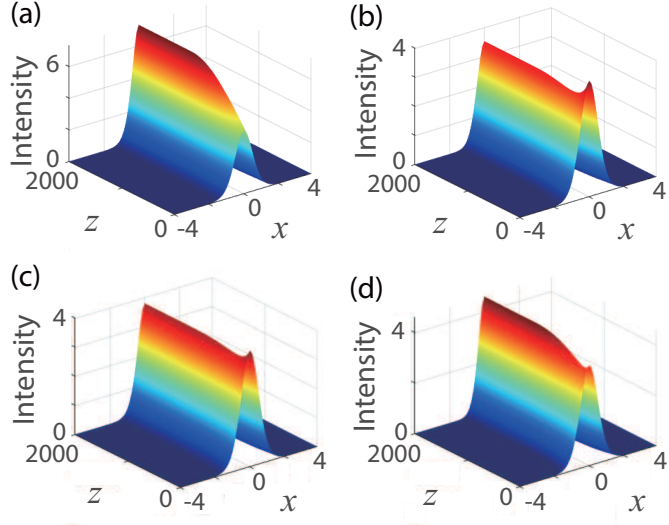}}}
\end{center}
\par
\vspace{-0.25in}
\caption{{\protect\small The application of the adiabatic transformation (%
\protect\ref{excite-s}) to the nonlinear mode (\protect\ref{nlm1}) with
initial parameters $\protect\alpha _{1}=2,\left( \protect\phi _{0}\right)
_{1}=2,\left( w_{0}\right) _{1}=3$: (a) }$\left( {\protect\small w_{0}}%
\right) _{1}=${\protect\small $3\rightarrow \left( w_{0}\right) _{2}=5$; (b)
$\protect\alpha _{1}=2\rightarrow \protect\alpha _{2}=3$; (c) }$\left(
{\protect\small \protect\phi _{0}}\right) _{1}${\protect\small $%
=2\rightarrow \left( \protect\phi _{0} \right) _{2}=1$; (d) the joint action of
the excitations in (a), (b) and (c). }}
\label{exc}
\end{figure}

\subsubsection*{3.1.3 \thinspace\ The adiabatic excitations of nonlinear
modes}

Here we address the case when the central position of the GS-II potential (%
\ref{GS}) is fixed at the origin, aiming to generate additional stable
solitons from input given by Eq. (\ref{nlm1}) by varying parameters of the
potential well adiabatically as functions of the propagation distance,
namely, $\alpha \rightarrow \alpha (z)$, $\phi _{0}\rightarrow \phi _{0}(z)$
or $w_{0}\rightarrow w_{0}(z)$, cf. Ref.~\cite{yan2015solitons}.
Accordingly, Eq. (\ref{nls}) with the $\mathcal{PT}$-symmetric GS-II
potential (\ref{GS}) is replaced  by
\begin{equation}
i\frac{\partial \psi }{\partial z}+\left[ \frac{\partial ^{2}}{\partial {%
x^{2}}}+V(x,z)+iW(x,z)+|\psi |^{2}\right] \psi=0 ,  \label{tnls}
\end{equation}%
where the $z$-dependent potentials $V(x,z)\,$and $W(x,z)$ are given by Eqs.~(%
\ref{GS}) and (\ref{AB}) with $\alpha \rightarrow \alpha (z)$, $\phi
_{0}\rightarrow \phi _{0}(z)$ and $w_{0}\rightarrow w_{0}(z)$.
Here, three scenarios of the adiabatic variation of parameters $\alpha (z)$,
$\phi _{0}(z)$ and $W_{0}(z)$ are chosen in the form of%
\begin{equation}
\!\Xi (z)=\!%
\begin{cases}
(\Xi _{2}\!-\!\Xi _{1})\!\sin \!\left( \dfrac{\pi z}{z_{\max }}\!\right)
\!+\!\Xi _{1}, & \text{$0\leq z<$}z_{\max }/2, \\
\Xi _{2},\, & \text{$z\geq $}z_{\max }/2,%
\end{cases}
\label{excite-s}
\end{equation}%
where 
$\Xi _{1,2}$ denote the initial and final values.

First, we apply the single-parameter variation to the bright soliton (\ref%
{nlm1}), which is taken as the initial condition, and its propagation is
governed by Eq.~(\ref{tnls}). When $w_{0}(z)$ is replaced with $\Xi (z)$
given by Eq.~(\ref{excite-s}), whereas $\alpha (z)\equiv \alpha $ and $\phi
_{0}(z)\equiv \phi _{0}$ are fixed, we can adiabatically transform the
initially stable localized mode produced by Eq.~(\ref{nlm1}) at $(\alpha
,\phi _{0},\left( w_{0}\right) _{1})=(2,2,3)$ to another stable mode
corresponding to parameters $(\alpha ,\phi _{0},\left( w_{0}\right)
_{2})=(2,2,5)$, see Fig.~\ref{exc}(a), even if the corresponding linear $%
\mathcal{PT}$-symmetry is broken [see Fig.~\ref{spe1}(b)] and the
corresponding exact soliton (\ref{nlm1}) is unstable [see Fig.~\ref{exa1}%
(c)]. This is an unexpected result, since the stability of the transformed
soliton should in general remain consistent with that of the exact soliton (%
\ref{nlm1}) for the same parameters.

Replacing parameter $\alpha (z)$ or $\phi _{0}(z)$ by expression Eq.~(\ref%
{excite-s}) and keeping the other parameters unchanged, similar
transformations of soliton (\ref{nlm1}) can be executed as well. As a
result, non-monotonous adiabatic-transform scenarios are observed in Figs.~%
\ref{exc}(b, c). It is worthy to note that in Fig.~\ref{exc}(c) we need to
make the value of $\left( \phi _{0}\right) _{2}$ smaller than $\left( \phi
_{0}\right) _{1}=2$, to produce the stable adiabatic transform. Otherwise,
the potential parameters in the final state, such as, e.g., $(\alpha ,\left(
\phi _{0}\right) _{2},w_{0})=(2,3,3)$, will make the soliton (\ref{nlm1})
unstable [see Fig.~\ref{exa1}(e)], destabilizing the entire procedure.

Next, the simulations demonstrate that the adiabatic transformation remains
stable if the substitution (\ref{excite-s}) is applied to all the three
parameters under the consideration, $(\alpha ,\phi _{0},w_{0})$. In this
case, Fig.~\ref{exc}(d) shows that the result of the simultaneous variation
of the three parameters seems as a superposition of the results produced by
the single-parameter variations.
Thus, the exact soliton (\ref{nlm1}) can be stably transformed to another
nonlinear mode by the properly applied adiabatic variation of the
potential's parameters.

\begin{figure}[t]
\begin{center}
\vspace{0.05in} \hspace{-0.05in}{\scalebox{0.6}[0.5]{%
\includegraphics{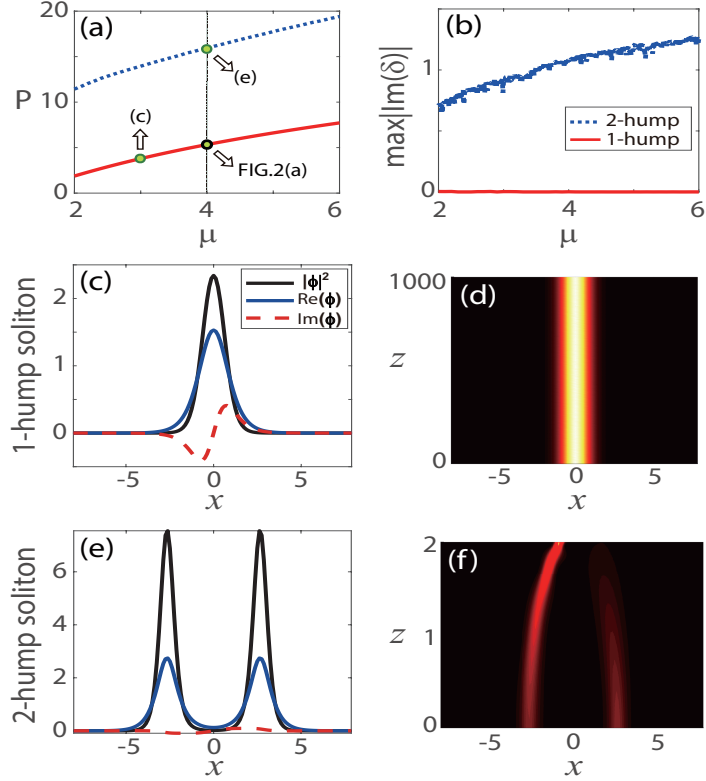}}}
\end{center}
\par
\vspace{-0.25in}
\caption{{\protect\small (a) The power and (b) the corresponding
linear-stability versus the propagation constant $\protect\mu $, for the
fundamental ($1$-hump) and first-excited-state ($2$-hump) solitons, produced
by numerical methods. The numerically found solitons and their propagation:
(c,d) for $\protect\mu =3$ (fundamental soliton), (e,f) $\protect\mu =4$
(first-excited-state soliton). Other parameters are $g=1,\protect\alpha %
=2,\protect\phi _{0}=2,w_{0}=3$. }}
\label{num1}
\end{figure}

\begin{figure*}[t]
\begin{center}
\vspace{0.05in} \hspace{-0.05in}{\scalebox{0.45}[0.45]{%
\includegraphics{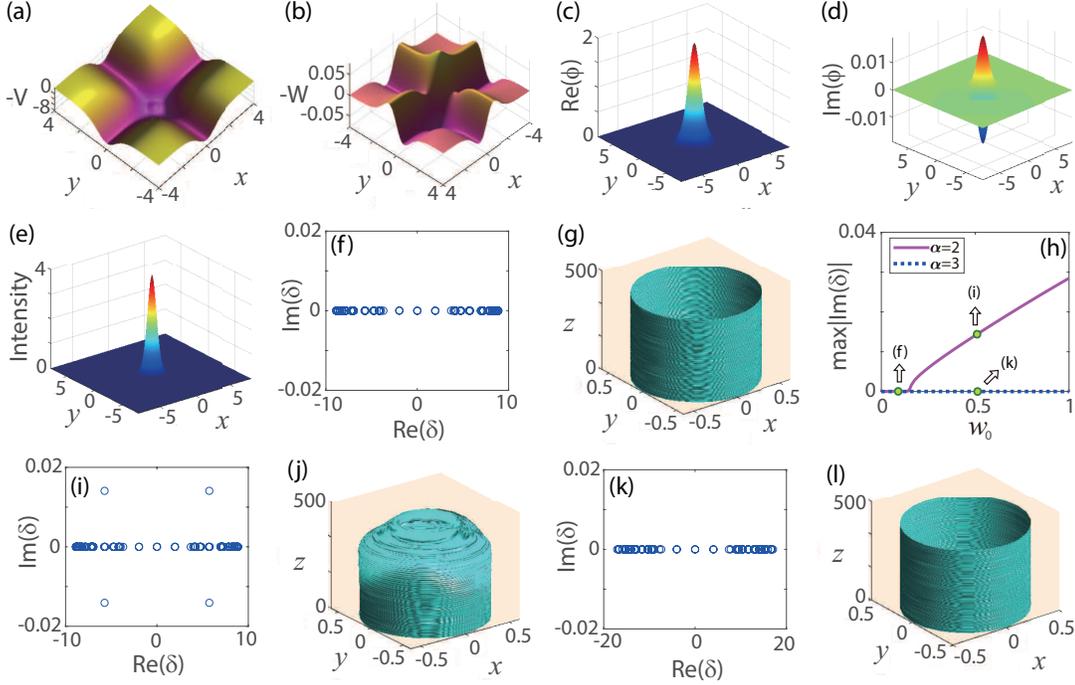}}}
\end{center}
\par
\vspace{-0.15in}
\caption{{\protect\small (color online). (a,b) Real and imaginary components
of the 2D $\mathcal{PT}$-symmetric generalized Scarf-II potential, $%
U=-V(x,y)-iW(x,y)$, with $(\protect\alpha ,\protect\phi _{0},w_{0})=(2,2,0.1)
$, (c,d,e) Real and imaginary parts and intensity of the corresponding 2D
exact soliton. (f) The linear stability spectrum of the soliton. (g) The
stable propagation. (h) The change of the linear-stability index as a
function of $w_{0}$, at $\protect\alpha =2\ $and$~3$. The linear stability
spectrum and propagation of the 2D exact solitons are displayed at $\protect%
\alpha =2,w_{0}=0.5$ (i,j); $\protect\alpha =3,w_{0}=0.5$ (k,l). Other
parameters are $g=1$.}}
\label{sol2}
\end{figure*}

\subsubsection*{3.1.4 \thinspace\ Internal modes and high-order solitons}

The preceding analysis is focused on exact solitons (\ref{nlm1}) of Eq.~(\ref%
{stat}) with $\mu =\alpha ^{2}$. It is, however, natural to expect that Eq.~(%
\ref{stat}) may produce other modes for different values of $\mu $. They can
be produced by means of the iteration scheme developed in Sec. 2.3. To this
end, we choose a Gaussian input,
\begin{equation}
\phi _{\text{ini}}(x)=\rho \,|x|^{m}\,e^{-w^{-2}x^{2}},\quad m\in \mathbb{Z}%
^{+}  \label{ic}
\end{equation}%
with real amplitude $\rho $ and width $w$. The iteration scheme (\ref{ite})
demonstrates that the initial Gaussian (\ref{ic}) converges to the exact
soliton (\ref{nlm1}) with $\mu =\alpha ^{2}$. For instance, at $\alpha =2$,
hence $\mu =4$, we numerically obtain a localized nonlinear mode which is
identical to the exact soliton (\ref{nlm1}), marked by the black circle in
Fig.~\ref{num1}(a). For values of $\mu $ close to $4$, with fixed $\alpha =2$%
, we can take soliton (\ref{nlm1}) at $\mu =4$ as the input, so as to
improve the convergence speed of the numerical scheme and produce a family
of solitons (fundamental modes) shown by the solid red line in Fig.~\ref%
{num1}(a), as a function of $\mu $. At $\mu =3$, a representative
fundamental mode is exhibited in Fig.~\ref{num1}(c). The wave propagation is
stable in a long time, see Fig.~\ref{num1}(d).

On the other hand, we can produce families of higher-order nonlinear modes
by means of the modified squared-operator iteration method~\cite%
{yang2007universally}. In particular, the generated family of the first
excited states is plotted by the blue dotted line in Fig.~\ref{num1}(a). A
typical soliton representing the first excited state (see Fig.~\ref{num1}%
(e)) turns out to be unstable in direct simulations (see Fig.~\ref{num1}(f)).

Linear-stability plots for the ground and first excited states are presented
in Fig.~\ref{num1}(b). It is seen that the ground states are completely
stable, however the first excited states are unstable, in agreement with
Figs.~\ref{num1}(d,f). It has been checked that excited states of higher
orders are subject to a still stronger instability.

\subsection*{3.2\thinspace\ 2D solitons and stability}

\subsubsection*{3.2.1 \thinspace\ Exact spatial solitons}

In this section, we address the formation of 2D spatial solitons and
their stability in the 2D geometry determined by the $\mathcal{PT}$%
-symmetric GS-II potential defined by Eqs.~(\ref{GS-2D}) and~(\ref{AB12})
(see Figs.~\ref{sol2}(a,b)). In this case, Eq.~(\ref{nls}) becomes
\begin{equation}
i\frac{\partial \psi }{\partial z}+\nabla _{2}^{2}\psi
\!+\![V(x,y)+iW(x,y)]\psi +g|\psi |^{2}\psi =0,  \label{nls2D}
\end{equation}%
with $\nabla _{2}=(\partial _{x},\partial _{y})$. 
Then, an exact localized solution of Eq.~(\ref{nls2D}) is sought for as
\begin{equation}  \label{stasol2}
\psi=\phi _{0}(\mathrm{sech}x\,\mathrm{sech}y)^{\alpha }\exp\!\left(\!2i\alpha ^{2}z\!\!+\!%
\dfrac{iw_{0}}{3\alpha }\!\!\!\mathop{\sum}\limits_{x_{j}=x,y}
\int_{0}^{x_{j}}\!\!\mathrm{sech}^{\alpha }sds\!\right) .
\end{equation}

The stationary solution $\phi (x,y)$ is exhibited by its real and imaginary
components, as well as intensity $|\phi (x,y)|^{2}$, in Figs.~\ref{sol2}%
(c,d,e). The linear-stability spectrum without any imaginary parts in Fig.~%
\ref{sol2}(f) reveals that this stationary soliton is stable. Direct
propagation simulations of this 2D soliton further confirm its stability, as
shown by contour lines of the soliton intensity, $\left( 1/2\right) \max
(|\psi (x,y,0)|^{2})$, in Fig.~\ref{sol2}(g). Moreover, the solid curve in
Fig.~\ref{sol2}(h) shows that the 2D modes are linearly stable only below
the critical value, i.e., at $w_{0}<\left( w_{0}\right) _{\mathrm{crit}%
}\approx 0.15$, while other parameters are fixed. For instance, the
above-mentioned 2D soliton (\ref{stasol2}) is indeed stable, as it was found
for $w_{0}=0.1<\left( w_{0}\right) _{\mathrm{crit}}$, while the increase of $%
w_{0}$ to $0.5$ makes the 2D soliton unstable, as shown by its
linear-stability spectrum and nonlinear propagation, see Figs.~\ref{sol2}%
(i,j). However, the blue dotted line in Fig.~\ref{sol2}(h) shows that the
increase of $\alpha $ to $3$ makes the 2D modes linearly stable in a broader
range of $w_{0}$, as is corroborated by Figs.~\ref{sol2}(k,l).

\begin{figure}[t]
\begin{center}
\vspace{0.05in} \hspace{-0.05in}{\scalebox{0.56}[0.56]{%
\includegraphics{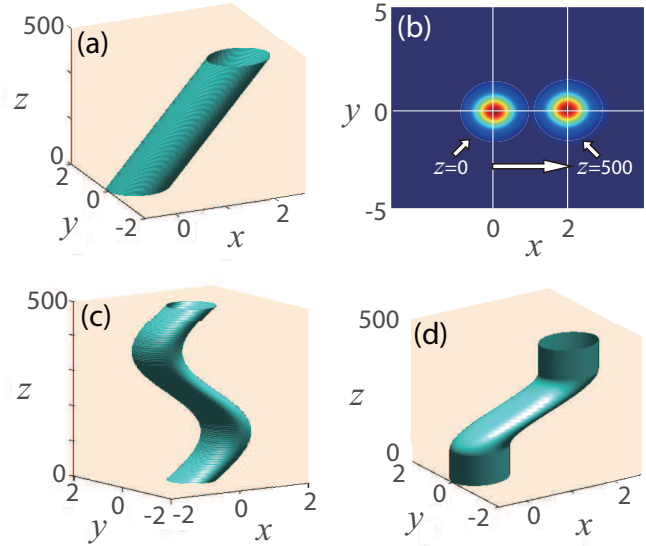}}}
\end{center}
\par
\vspace{-0.2in}
\caption{{\protect\small Evolution patterns of the 2D soliton (\protect\ref%
{stasol2}) for different modulation patterns of the GS-II potential (\protect
\ref{GS-2D}) with $x_{j}-\tilde{x}_{j0}(z)$: (a) $\tilde{x}%
_{10}(z)=2z/z_{\max }$; (b) the respective initial and final states; (c) $%
\tilde{x}_{10}(z)=\sin (2\protect\pi z/z_{\max })$; (d) for $\tilde{x}%
_{10}(z)$ given by Eq.~(\protect\ref{mpf}). Other parameters are $g=1,\,%
\protect\alpha =\protect\phi _{0}=2,\,w_{0}=0.1$. }}
\label{pat2}
\end{figure}

\subsubsection*{3.2.2 \thinspace\ The evolution of 2D spatial solitons under
perturbations and excitation}

When the 2D GS-II potential well (\ref{GS-2D}) varies along the propagation
distance, the governing 2D model becomes
\begin{equation}
i\frac{\partial \psi }{\partial z}+\left[\nabla_{2}^{2}+V(x,y,z)+iW(x,y,z)+|\psi |^{2}\right]\psi=0.
\label{tnls2D}
\end{equation}%
%
%
%

Like the 1D case, the first issue to address is the possibility of stable
propagation of the soliton initiated by solution (\ref{stasol2})\ when the
central position of the 2D GS-II potential (\ref{GS-2D}) depends on the
propagation distance, that is, we considered the potential $V(x,z)$ and $%
W(x,z)$ given by Eq.~(\ref{GS-2D}) with $x_{j}-\tilde{x}_{j0}(z)$,
where
we set $\tilde{x}_{j0}(0)=0$ to make the 2D nonlinear mode (\ref{stasol2}) a
correct solution of Eq.~(\ref{tnls2D}) under the GS-II potential (\ref{GS-2D}%
) at $z=0$.

Here, we always set $\tilde{x}_{20}(z)=0$ and consider three options for $%
\tilde{x}_{10}(z)$, as follows:

\begin{itemize}
\item The linear perturbation: When the central position of the 2D GS-II
potential (\ref{GS-2D}) with $x_{j}-\tilde{x}_{j0}(z)$ uniformly moves to
the right in the $x$ direction with the growth of the propagation distance,
such as $\tilde{x}_{10}(z)=2z/z_{\max }$, the soliton propagates stably
along the same $x$ direction, as is shown in Fig.~\ref{pat2}(a). In
particular, Fig.~\ref{pat2}(b) clearly indicates the central position of the
2D soliton in the initial and final states.

\item The periodic perturbation: When the central position of the 2D GS-II
potential (\ref{GS-2D}) with $x_{j}-\tilde{x}_{j0}(z)$ oscillates
periodically in the $x$ direction as a function of the propagation distance,
i.e., $\tilde{x}_{10}(z)=\sin (2\pi z/z_{\max })$, the soliton initiated by
Eq. (\ref{stasol2}) demonstrates stable snaking propagation, see Fig.~\ref%
{pat2}(c). In this case, the final central position of the 2D soliton is the
same as the initial one, both being at the origin. The same result is true
for the periodic modulation with the double frequency, $\tilde{x}%
_{10}(z)=\sin (4\pi z/z_{\max })$.

\item A mixed perturbation: When the path $\tilde{x}_{10}(z)$ is taken as
the following piecewise function.
\begin{equation}
\tilde{x}_{10}(z)\!=\!%
\begin{cases}
0, & \!\!\text{$0\leq z\leq \dfrac{z_{\max }}{4}$},\vspace{0.1in} \\
\sin \!\left( \dfrac{2\pi z}{z_{\max }}\!-\!\pi \right) \!+\!1, & \!\!\text{$%
\dfrac{z_{\max }}{4}<z\leq \dfrac{3z_{\max }}{4}$},\vspace{0.1in} \\
2, & \!\!\text{$\dfrac{3z_{\max }}{4}<z\leq z_{\max }$},%
\end{cases}
\label{mpf}
\end{equation}%
the soliton also moves stably, staying trapped in the potential well (\ref%
{GS-2D}) with $x_{j}-\tilde{x}_{j0}(z)$, in spite of the presence of
inflection points, see Fig. \ref{pat2}(d).
\end{itemize}

Thus, similar to the 1D case, the stable 2D nonlinear mode (\ref{stasol2})
propagates in the robust form along the trajectory determined by the central
position of potential well (\ref{GS-2D}) with $x_j-\tilde{x}_{j0}(z)$.

\begin{figure}[t]
\begin{center}
\vspace{0.05in} \hspace{-0.05in}{\scalebox{0.5}[0.5]{%
\includegraphics{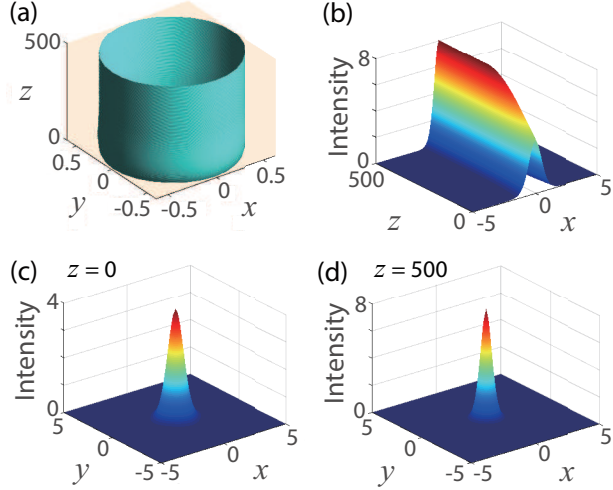}}}
\end{center}
\par
\vspace{-0.2in}
\caption{{\protect\small The transformation of the 2D nonlinear mode (%
\protect\ref{stasol2}) in Eq.~(\protect\ref{tnls2D}) with the potential well
(\protect\ref{GS-2D}) in which $\protect\alpha \rightarrow \protect\alpha (z)
$, $\protect\phi _{0}\rightarrow \protect\phi _{0}(z)$ and/or $%
w_{0}\rightarrow w_{0}(z)$: (a) the contour plot of the intensity of the
nonlinear mode; (b) the corresponding propagation in the $y=0$ cross
section; (c) the initial-state soliton; (d) the final-state soliton. Here $%
\protect\alpha (z)$ is given by Eq.~(\protect\ref{excite-s2D}) and other
parameters are $g=1,\protect\phi _{0}=2,w_{0}=0.5$. }}
\label{exc2}
\end{figure}

Another issue is whether is it possible to transform the stable 2D soliton
by adiabatically altering the potential's parameters. For this purpose,
similar to the 1D case, we make the potential-well parameters $(\alpha ,\phi
_{0},w_{0})$ in Eq.~(\ref{GS-2D}) varying, i.e., $\alpha \rightarrow \alpha
(z)$, $\phi _{0}\rightarrow \phi _{0}(z)$ and/or $w_{0}\rightarrow w_{0}(z)$%
. 

\begin{figure*}[t]
\begin{center}
\vspace{0.05in} \hspace{-0.05in}{\scalebox{0.5}[0.5]{%
\includegraphics{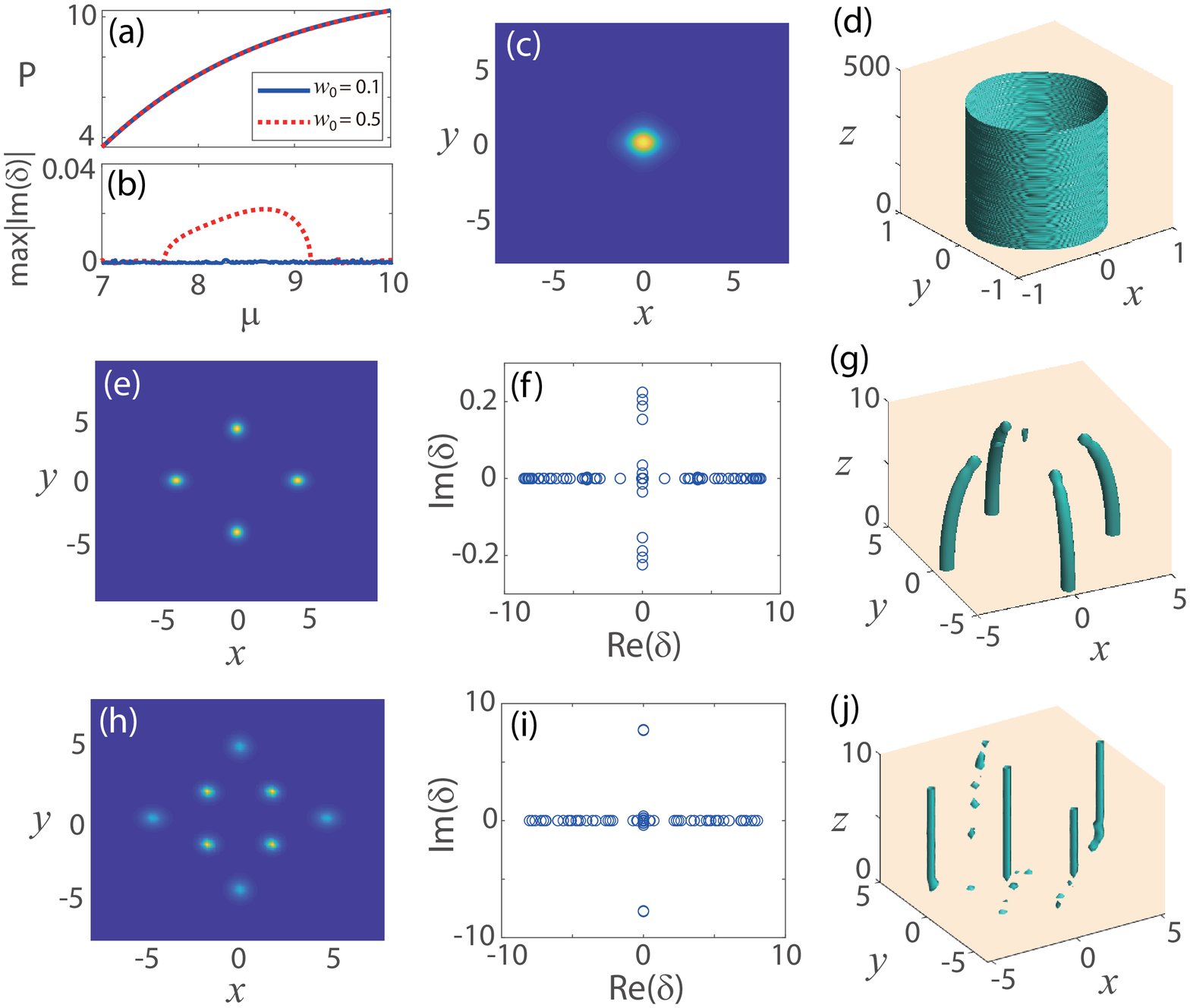}}}
\end{center}
\par
\vspace{-0.15in}
\caption{{\protect\small (color online). (a) The power and (b) the
corresponding linear-stability eigenvalue of the 2D fundamental solitons
versus the propagation constant $\protect\mu $ for $w_{0}=0.1$ and $0.5$,
produced by the numerical solution. (c) The numerically found fundamental
soliton depicted by means of the intensity, and (d) its propagation with $%
\protect\mu =7$, nearby the exact solution corresponding to propagation
constant $\protect\mu =8$. Higher-order vortex solitons, linear-stability
spectra, and the transmission dynamics with $w_{0}=0.1$: (e, f, g) $m=1$
(4-vortex), (h, i, j) $m=2$ (8-vortex). Other parameters are $g=1,\protect%
\alpha =\protect\phi _{0}=2$.}}
\label{num2}
\end{figure*}

Here we consider the case of the switch of $\alpha (z)$ in the form of Eq.~(%
\ref{excite-s}) and $\phi _{0}(z)\equiv \phi _{0},\,w_{0}(z)\equiv w_{0}$.
More specifically, taking $\phi _{0}(z)\equiv 2$, $w_{0}(z)\equiv 0.5$, and
\begin{equation}
\!\alpha (z)=\!%
\begin{cases}
\sin \left( \pi z/z_{\max }\right) +2, & \text{$0\!\leq z<z_{\max }/2$},%
\vspace{0.1in} \\
3,\, & \text{$z\geq z_{\max }/2,$}%
\end{cases}
\label{excite-s2D}
\end{equation}%
we start with an initially unstable 2D soliton (\ref{stasol2}) at $(\alpha
,\phi _{0},w_{0})=(2,2,0.5)$, transforming it into another \emph{stable}
localized mode with parameters $(\alpha ,\phi _{0},w_{0})=(3,2,0.5)$, as
shown by contours and profiles in Figs.~\ref{exc2}(a,b). Here, Fig.~\ref%
{exc2}(c) displays the initial unstable soliton [see Fig.~\ref{sol2}(j)],
while Fig.~\ref{exc2}(d) exhibits the stable final-state soliton. It is
worthy to note that the final-state soliton is a new nonlinear mode whose
shape (with the peak value $\phi _{\max }^{2}\approx 8$) is conspicuously
different from that of the exact 2D soliton (\ref{stasol2}) at $(\alpha
,\phi _{0},w_{0})=(3,2,0.5)$, whose peak intensity is $\phi _{\max }^{2}=4$.

Thus, the 2D exact soliton (\ref{stasol2}) can be stably transformed to
other nonlinear localized modes, like in the 1D case, by means of the
adiabatic variation of the parameters of the GS-II potential (\ref{GS-2D})
with $\alpha \rightarrow \alpha (z)$, $\phi _{0}\rightarrow \phi _{0}(z)$
and/or $w_{0}\rightarrow w_{0}(z)$.

\subsubsection*{3.2.3 \, 2D internal modes and high-order solitons}

Once the parameters of the 2D system (\ref{nls2D}) and GS-II potential (\ref%
{GS-2D}) with $\alpha \rightarrow \alpha (z)$, $\phi _{0}\rightarrow \phi
_{0}(z)$ and/or $w_{0}\rightarrow w_{0}(z)$ are fixed as per Eq.~(\ref%
{stasol2}), we obtain the single exact fundamental mode corresponding to $%
\mu =2\alpha ^{2}$. Varying parameter $\mu $, other nonlinear localized
modes can be found, including 2D fundamental solitons.

To address this possibility, we start with the potential's parameters $%
(\alpha ,\phi _{0},w_{0})=(2,2,0.1)$. By means of the numerical iteration
technique developed in Sec. 2.3 and taking the exact 2D soliton at $\mu =8$
as the input, a family of new 2D fundamental solitons can be found with
values of the propagation constant around $\mu =8$, as shown by the solid
blue line in Fig.~\ref{num2}(a). Figure~\ref{num2}(a) also plots the power
curve of the 2D fundamental solitons, with $w_{0}=0.5$, by the red dotted
line, which is the same as the one of the 2D fundamental solitons with $%
w_{0}=0.1$. This result is reasonable because, at least for $\mu =8$, we see
from Eq.~(\ref{stasol2}) that the exact 2D fundamental solitons do not
change their density as $w_{0}$ varies.

\begin{figure*}[t]
\begin{center}
\vspace{0.05in} \hspace{-0.05in}{\scalebox{0.45}[0.45]{%
\includegraphics{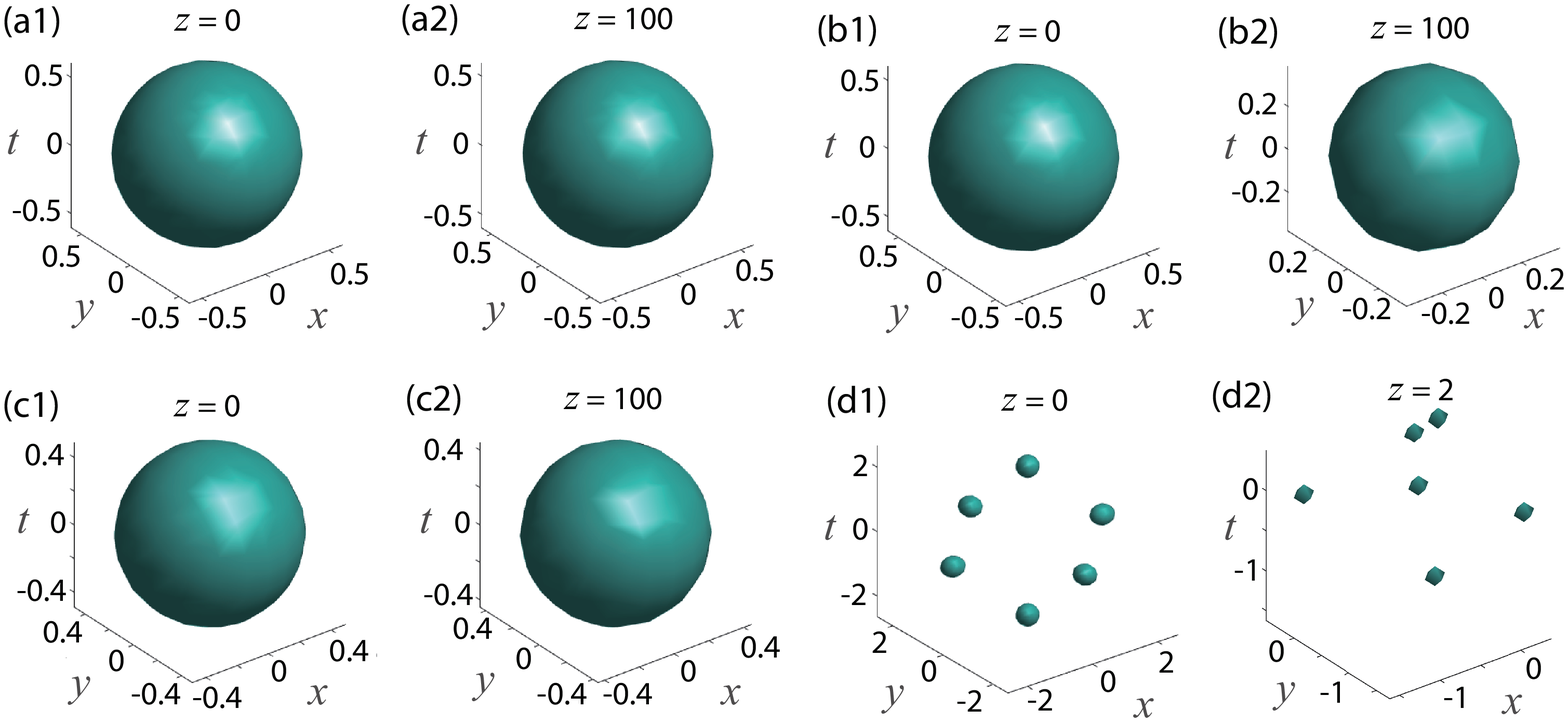}}}
\end{center}
\par
\vspace{-0.15in}
\caption{{\protect\small (color online). The isosurface evolution of 3D
exact solitons (\textquotedblleft light bullets") with $\protect\mu =3%
\protect\alpha ^{2}$: (a1, a2) $\protect\alpha =2,w_{0}=0.1$; (b1, b2) $%
\protect\alpha =2,w_{0}=0.5$; (c1, c2) $\protect\alpha =3,w_{0}=0.5$.
(d1,d2) The isosurface evolution of a higher-order \textquotedblleft bullet"
obtained numerically with the same parameters as in (a1,a2). Other
parameters are $g=1,\, \protect\phi _{0}=2$. }}
\label{sol3}
\end{figure*}

However, the linear-stability analysis, as shown in Fig.~\ref{num2}(b),
reveals that the stability of these two families of solitons is very
different. The first family, with $w_{0}=0.1$, is entirely stable, while the
family with $w_{0}=0.5$ is partially unstable, especially near $\mu =8$.
This happens because the growth of $w_{0}$ increases the strength of the
gain-loss distribution, which easily leads to instability of the soliton.
Although the family of solitons with $w_{0}=0.5$ is unstable near $\mu =8$,
we can find a series of stable modes from Fig.~\ref{num2}(b) by adjusting
the propagation constant appropriately. For example, when $\mu =7$, a 2D
fundamental mode is numerically obtained and the propagation simulation
confirms that it is stable [see Figs.~\ref{num2}(c) and (d)], which is
consistent with the result of the linear-stability analysis in Fig.~\ref%
{num2}(b).

In addition, we have found that families of 2D vortex solitons exist too,
but they are unstable. For instance, using the modified squared-operator
iteration method with the input of the following form,
\begin{equation}
\phi _{\text{ini}}(x,y)=\rho \,(x^{2}+y^{2})^{m/2}\,\exp \left( -\frac{x^{2}+y^{2}%
}{w^{2}}\right) ,\quad  m\in \mathbb{Z}^{+},  \label{nic}
\end{equation}%
where amplitude $\rho $ and width $w$ need to be chosen appropriately, we
can produce four-peak and eight-peak vortex solitons [see Figs.~\ref{num2}%
(e, h)], with winding numbers $m=1$ and $2$. The computation of their
linear-stability spectrum and direct simulations clearly demonstrate that
they are unstable solutions, see Figs.~\ref{num2}(f)-(j).

\subsection*{3.3\thinspace\ 3D \textquotedblleft light bullets" and their
stability}

\subsubsection*{3.3.1 \thinspace\ 3D \textquotedblleft light bullets"}

3D spatiotemporal solitons, alias \textquotedblleft light bullets", are also
a subject of great significance in nonlinear optics \cite%
{malomed2016multi,kartashov2019front}. In this section we address the
formation of 3D solitons in the 3D $\mathcal{PT}$-symmetric GS-II potential.
The 3D governing equation with this potential and Kerr nonlinearity is
written as~\cite{dai2014stable}
\begin{equation}
i\frac{\partial \psi }{\partial z}+\left[ \nabla
_{3}^{2}+V(x,y,t)+iW(x,y,t)+g|\psi |^{2}\right] \psi =0,  \label{nls3D}
\end{equation}%
with $\nabla _{3}=(\partial _{x},\partial _{y},\partial _{t})$,\thinspace\ $%
V(x,y,t),\,W(x,y,t)$ given by Eqs.~(\ref{GS-3D}) and (\ref{AB12}). 
A necessary caveat is that, unlike the 1D and 2D setting, the experimental
realization of fully three-dimensional potentials in optics is quite
difficult, as creating a particular potential form in the temporal direction
is a challenging problem. For this reason, the results presented below for
3D solutions have a methodological interest, rather than promising direct
physical implementation. On the other hand, they may be directly interpreted
in terms of the Gross-Pitaevskii equation for Bose-Einstein condensates,
subject to the action of the 3D $\mathcal{PT}$-symmetric potential (the same
is possible for the 1D and 2D settings considered above)~\cite%
{dast2013eigenvalue}.

In the 3D $\mathcal{PT}$-symmetric GS-II potential (\ref{GS-3D}), the exact
3D solution of Eq.~(\ref{nls3D}) can be obtained in the stationary form,
\begin{equation}\label{stasol3}
\begin{array}{rl}
\psi=&\d\phi _{0}(\mathrm{sech}x\,\mathrm{sech}y\,\mathrm{sech}t)^{\alpha }\exp\left(3i\alpha ^{2}z+\dfrac{iw_0}{3\alpha}\mathop{\sum}\limits_{x_{j}=x,y,t}%
\d\int_{0}^{x_{j}}\mathrm{sech}^{\alpha }sds\right),%
\end{array}%
\end{equation}

In particular, for $\alpha =1$ and $\phi _{0}=\sqrt{(w_{0}^{2}\!-\!9v_{0}\!+%
\!18)/(9g)}$ 
the 3D $\mathcal{PT}$-symmetric GS-II potential (\ref{GS-3D}) reduces to
\begin{equation}
\begin{array}{l}
V=(w_{0}^{2}/9+2)\!\!\mathop{\sum}\limits_{x_{j}=x,y,t}\mathrm{sech}^{2}x_{j}
+(v_{0}^{2}\!-\!2\!-\!w_{0}^{2}/9)\!\!\mathop{\prod}\limits%
_{x_{j}=x,y,t}\mathrm{sech}^{2}x_{j},\quad
W\!=\!w_{0}\mathop{\sum}\limits_{x_{j}=x,y,t}\mathrm{sech}x_{j}\tanh x_{j}.%
\end{array}
\label{GS-3D11}
\end{equation}%
and solution (\ref{stasol3}) becomes
\begin{equation}
\psi =\phi _{0}\left( \mathrm{sech}x\right) \,\left( \mathrm{sech}y\right)
\vspace{3pt}\left( \mathrm{sech}t\right) \,e^{i\varphi (x,y,t)+3iz},
\end{equation}%
with $\varphi (x,y,t)=\frac{w_{0}}{3}\mathop{\sum}\limits_{x_{j}=x,y,t}%
\arctan (\sinh x_{j})$, whose stability has been investigated~\cite%
{dai2014stable}.

However, for $\alpha >0$ and $\alpha \neq 1$, the 3D $\mathcal{PT}$%
-symmetric GS-II potential (\ref{GS-3D}) is essentially more general than
the one with $\alpha =1$, which makes the corresponding 3D solution
significantly more general too. Here we take $\alpha =2$ as an example to
illustrate the stability of the analytical 3D solution (\ref{stasol3}).
Fig.~\ref{sol3}(a1) displays the isosurface plot of a typical 3D soliton
with $(\phi _{0},w_{0})=(2,0.1)$. Just as in the 2D case, the soliton
exhibits stable propagation in the course of long simulations [see Fig.~\ref%
{sol3}(a2)], while increasing $w_{0}$ to $0.5$ makes the soliton compressed
after travelling a certain distance, as shown in Figs.~\ref{sol3}(b1) and %
\ref{sol3}(b2). In this case, if the value of $\alpha $ increases properly,
the soliton again keeps its waveform unchanged and demonstrates stable
propagation, see Figs.~\ref{sol3}(c1) and \ref{sol3}(c2).

We have also numerically investigated higher-order 3D solitons, finding that
they are unstable. For instance, a higher-order six-peak soliton, produced
by the numerical solution for $(\phi _{0},w_{0})=(2,0.1)$, is displayed in
Fig.~\ref{sol3}(d1). However, it is strongly unstable and quickly collapses
in the course of the propagation, see Figs.~\ref{sol3}(d2).

\subsubsection*{3.3.2 \thinspace\ $n$-dimensional nonlinear localized mode}

Finally, we further generalize the $\mathcal{PT}$-symmetric GS-II potential (%
\ref{GS}) to an abstract $n$-dimensional form,
\begin{equation}
\begin{aligned} & V({\bm x})\!=\!\!\sum_{j=1}^{n}(v_1{\rm sech}^2
x_j\!+\!v_{21}{\rm sech}^{2\alpha}x_{j})\!+\!v_{22}\!\prod_{j=1}^{n}{\rm
sech}^{2\alpha}x_j,\quad
 W({\bm x})\!=w_0\sum_{j=1}^{n} {\rm sech}^\alpha
x_j\tanh x_j, \end{aligned}  \label{MGS}
\end{equation}%
where $v_1$, $v_{21}$ and $v_{22}$ are subject, as above, to constraint (\ref%
{AB12}). By means of the analysis similar to that developed above in
subsection Sec. (2.1), one can readily find that Eq.~(\ref{nls}) admits a
particular exact soliton solution of the following stationary form:
\begin{equation}
\psi\!=\!\phi _{0}\!\prod_{j=1}^{n}\mathrm{sech}^{\alpha
}x_j\!\exp\!\left(\!in\alpha^2z\! \!+\!\frac{iw_{0}}{3\alpha }%
\sum_{j=1}^{n}\!\int_{0}^{x_j}\!\!\mathrm{sech}^{\alpha }s\,ds\right).
\label{nlmn}
\end{equation}%
With the help of the aforementioned numerical method, one can verify that it
is indeed a correct $n$-dimensional solution.

\begin{figure}[!t]
\begin{center}
\vspace{0.05in} \hspace{-0.05in}{\scalebox{0.45}[0.45]{%
\includegraphics{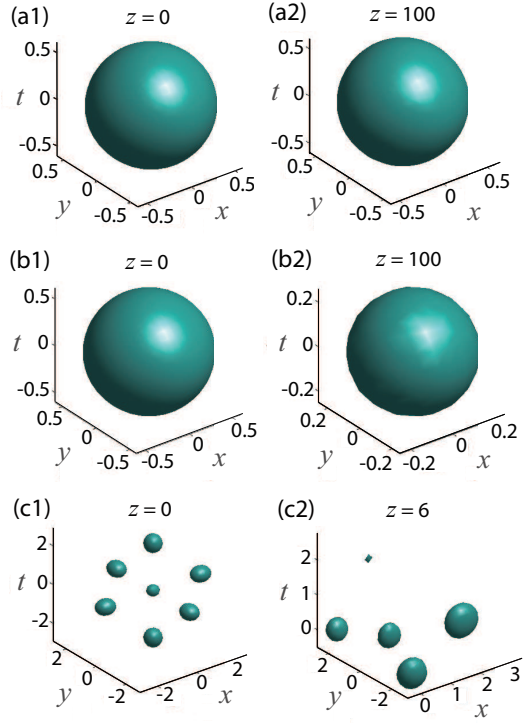}}}
\end{center}
\par
\vspace{-0.25in}
\caption{{\protect\small (color online). The isosurface evolution of 3D
light bullets with $\protect\mu =12$: (a1, a2) $w_{0}=0.1$, (b1, b2) $w_{0}=0.5$,
(c1, c2) The isosurface evolution of a 3D soliton with $\protect\mu =10$ and
$w_{0}=0.1$. Other parameters are $g=1,\, \protect\alpha =\protect\phi _{0}=2$.}
}
\label{sol3g1}
\end{figure}

\begin{figure}[!t]
\begin{center}
\vspace{-0.15in} \hspace{0.15in}{\scalebox{0.45}[0.45]{%
\includegraphics{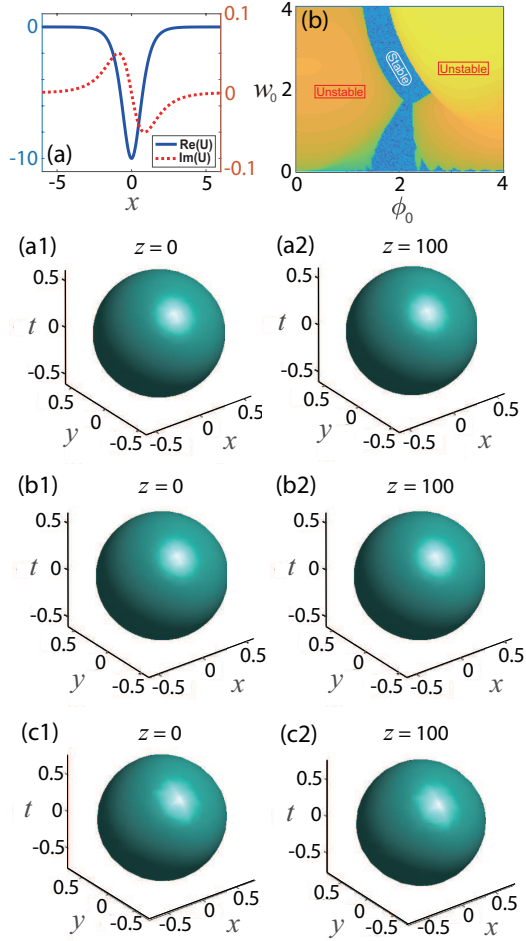}}}\hspace{-0.15in}
\end{center}
\par
\vspace{-0.25in}
\caption{{\protect\small (color online). 3D solutions under the
action of the self-defocusing nonlinearity, }$g=-1${\protect\small .}
{\protect\small (a) Real and imaginary parts of the $\mathcal{PT}$-symmetric
GS-II potential $U=-V(x)-iW(x)$ with $w_{0}=0.1$. (b) The linear-stability
map of exact solitons (\protect\ref{nlm1}) in the $(\protect\phi _{0},w_{0})$
 plane. The isosurface evolution of 3D exact solutions with $%
\protect\mu =12$: (a1,a2) $w_{0}=0.1$; (b1,b2) $w_{0}=0.5$. (c1,c2): The
isosurface evolution of the 3D soliton with $\protect\mu =10$ and $w_{0}=0.1$%
. Other parameters are $\protect\alpha =\protect\phi _{0}=2$.}}
\label{sol3g-1}
\end{figure}

\section*{4.\thinspace\ Multi-dimensional solitons in an
alternative $\mathcal{PT}$-symmetric GS-II potential}

\subsection*{4.1 \thinspace\ The second $n$-dimensional GS-II potential and
an exact solution}

In this section, we briefly consider another form of a $\mathcal{PT}$%
-symmetric GS-II potential in $n$ dimensions and the corresponding solutions
for localized modes. To this end, the $n$-dimensional $\mathcal{PT}$%
-symmetric GS-II potential (\ref{MGS}) is replaced by the following form:
\begin{equation}
\begin{aligned}
V({\bm x})=v_1\sum_{j=1}^{n} {\rm sech}^2x_{j}
+v_2\prod_{j=1}^{n}{\rm sech}^{2\alpha}x_{j},\quad
 W({\bm
x})=w_0\sum_{j=1}^{n} {\rm sech} x_{j}\tanh x_{j}, \end{aligned}
\label{MGSn}
\end{equation}%
where $v_1$, $v_2$ and $w_{0}$ are real parameters. If one considers the
constraints of $v_1,\, v_2,\, w_0$ as
\begin{equation}
v_1=\alpha (\alpha +1)+w_{0}^{2}/(2\alpha +1)^{2},\quad v_2=-g\phi _{0}^{2},
\label{ABM}
\end{equation}%
one can readily find that Eq.~(\ref{nls}) possesses the exact stationary
solution:
\begin{equation} \label{Mnlmn}
\psi ({\bm x},z)=\phi _{0}e^{i\varphi ({\bm x})+in\alpha
^{2}z}\prod_{k=1}^{n}\mathrm{sech}^{\alpha }x_{k},  \quad
\varphi ({\bm x})=\frac{w_{0}}{2\alpha +1}\sum_{k=1}^{n}\tan^{-1}(\sinh
x_{k}).
\end{equation}%
where both $\alpha $ and $\phi _{0}$ are real potential parameters, as above.

If $\alpha =1$, the GS-II potential (\ref{MGSn}), subject to constraint~(\ref%
{ABM}), is tantamount to the above GS-II potential defined by Eqs. (\ref{MGS}%
) and~(\ref{AB12}). Accordingly, the exact nonlinear mode (\ref{Mnlmn}) is
tantamount to one given above by Eq. (\ref{nlmn}). On the other hand, the
present $n$-dimensional GS-II potential (\ref{MGSn}) and exact mode (\ref%
{Mnlmn}) generalize results reported in Ref. \cite{chen2020soliton}, where
1D and 2D spatial solitons and their stability were investigated in detail.
Next we aim to elaborate the formation and propagation of solitons in the 3D
$\mathcal{PT}$-symmetric GS-II potential (\ref{MGSn}) under the action of
self-focusing and defocusing cubic nonlinearity.

\subsection*{4.2 \thinspace\ Stability of the 3D light bullets}

In the 3D case ($n=3$), the exact solution can be generated by Eq.~(\ref%
{Mnlmn}) and stay stable within a certain range of the potential's
parameters. More explicitly, setting $(x_{1},x_{2},x_{3})=(x,y,t)$, the 3D
GS-II potential becomes
\begin{equation}
\begin{array}{l}
V\!=\![\alpha (\alpha +1)+w_{0}^{2}/(2\alpha +1)^{2}]\mathop{\sum}\limits%
_{x_{j}=x,y,t}\mathrm{sech}^{2}x_{j}-g\phi _{0}^{2}\mathop{\prod}\limits_{x_{j}=x,y,t}\mathrm{sech}%
^{2\alpha }x_{j},\vspace{5pt} \\
W\!=\!w_{0}\mathop{\sum}\limits_{x_{j}=x,y,t}\mathrm{sech}x_{j}\tanh x_{j}.%
\end{array}
\label{GS-3D2}
\end{equation}%
With this potential, Eq.~(\ref{Mnlmn}) yields the following exact solution:
\begin{equation} \label{nlm3D2}
\psi \!=\!\phi _{0}\left[ \left( \mathrm{sech}x\right) \left( \mathrm{sech}%
y\right) (\mathrm{sech}t)\right] ^{\alpha }\,e^{i\varphi (x,y,t)+3i\alpha
^{2}z},\quad
\varphi (x,y,t)=\dfrac{w_{0}}{2\alpha +1}\mathop{\sum}\limits%
_{x_{j}=x,y,t}\arctan (\sinh x_{j}).%
\end{equation}

For $w_{0}=0.1$, Fig.~\ref{sol3g1}(a1) displays a representative 3D exact
solution, which is depicted by means of its isosurface. The solution is
stable, keeping its profile in the course of the the propagation up to $%
z=100 $, see Fig.~\ref{sol3g1}(a2). However, for $w_{0}=0.5$ the exact
solution (\ref{nlm3D2}) is compressed and loses its stability for a long
propagation distance, as seen in Figs.~\ref{sol3g1}(b1,b2). Furthermore, in
a vicinity of the propagation constant corresponding to the exact 3D
solution, we have found 3D\ higher-order solitons in the numerical form, as
shown in Fig. \ref{sol3g1}(c1). However, they are subject to strong
instability. For example, a short propagation distance, $z=6$, is sufficient
for the destruction of the higher-order mode in Fig.~\ref{sol3g1}(c2), built
from the input
\begin{equation}
\phi _{\text{ini}}({\bm x})=\rho \,|{\bm x}|^{m}\,e^{-{\bm x}^{2}/w^{2}},\quad |{\bm x%
}|=\sqrt{\sum_{i=1}^{n}{x_{i}^{2}}}.  \label{nic}
\end{equation}

In the case of the self-defocusing, $g=-1$, we have also explored solutions
supported by the $\mathcal{PT}$-symmetric GS-II potential in the 1D and
multidimensional settings. Here we only produce a brief overview of the
results. What is essentially different from the case of the self-focusing, $%
g=1$, is that the GS-II potential gets back to the single-well shape,
similar to the usual Scarf-II potential, see Fig.~\ref{sol3g-1}(a). Then, it
is easy to prove that such a potential may have entirely real linear energy
spectra. It also admits exact and stable localized states, although in a
relatively narrow range of the potential's parameters [see Fig.~\ref{sol3g-1}%
(b) for the 1D case]. Here we consider 3D solutions as examples. Fixing $%
\phi _{0}=2$ and $w_{0}=0.1$, we find that the respective exact 3D localized
solution propagates stably for a long distance, see Figs.~\ref{sol3g-1}%
(a1,a2). Even if increasing $w_{0}$ to $0.5$, the fundamental mode can retain its shape
over a long propagation distance [see Figs.~%
\ref{sol3g-1}(b1,b2)]. Thus, the stability of such solutions is apparently
weaker than for their counterparts with $g=1$ (in the case of the
self-focusing). In addition, for values of the propagation constant around
the exact 3D solution, we found a family of stable fundamental 3D modes in
the numerical form. For instance, Fig.~\ref{sol3g-1}(c1) exhibits a 3D
numerically found fundamental mode with $\mu =10$, while Fig.~\ref{sol3g-1}%
(c2) confirms its stability, as tested by the direct simulations.

\section*{ 5. \thinspace\ Conclusions and discussions}

In this work we have constructed a class of non-Hermitian $\mathcal{PT}$%
-symmetric potentials in the $n$-dimensional geometry, which is a
significant generalization of the classical Scarf-II potentials. In the
linear regime, we have proven that the 1D GS (generalized-Scarf)-II
potentials may possess purely real spectra in a certain area of the space of
the potential's parameters, with $\mathcal{PT}$-symmetry breaking occurring
at borders of this area. By carefully adjusting the localization parameter,
the range of the potential's parameters corresponding to the completely real
spectra can be further expanded. Similar results are obtained for
multidimensional settings. In the cubic-nonlinear media, $n$-dimensional
nonlinear localized modes are obtained in the exact analytical form. The
numerically performed linear-stability analysis reveals that the exact
nonlinear modes may be stable in a certain range of the potential's
parameters. Properly increasing the localization parameter, one can enlarge
the stability domain for the soliton solutions. Overall, the stability of
the exact solutions deteriorates with the increase of dimension $n$ or
strength $w_{0}$ of the the gain-loss term (the imaginary part of the $%
\mathcal{PT}$-symmetric potential). Using numerical approaches, we have
constructed continuous families of fundamental and excited-state solitons.
The former ones tend to be stable, while the latter solutions are subject to
strong instability. These unstable modes include 1D multimodal solitons, 2D
solitary vortices, and 3D multi-peak solutions. By means of adiabatic
modulations of the GS-II potential, the transformation of the solutions into
exact solitons can be performed. We have also demonstrated that stable
solitons can always propagate stably, being trapped in a slowly moving
potential well, which may be used for various manipulations of the optical
solitons. Finally, we have presented another category of $n$-dimensional $%
\mathcal{PT}$-symmetric GS-II potentials and the corresponding analytical
solutions for localized 3D\ modes, which may be stable in the model with
both self-focusing and defocusing signs of the Kerr (cubic) nonlinearity.

Lastly, we note that analytical and numerical methods developed in the
present work can be applied to more general NLS-like equations, with other
types of $\mathcal{PT}$-symmetric potentials or more complex nonlinearities.
In 1D, they may be written as
\begin{eqnarray}\label{PT-GNLS}
i\frac{\partial \psi }{\partial z}=\sum_{k=1}^{m}c_{k}(x)p^{k}\psi
+[V(x)+iW(x)]\psi -g(x)\hat{N}[\psi ,\psi _{x},|\psi |^{2},(|\psi
|^{2})_{x},...],
\end{eqnarray}%
where $\psi \equiv \psi (x,z)$ is the complex-valued wave function, $%
p=-i\partial /\partial x$ is the momentum operator, $c_{k}(x)$ and $g(x)$
are real even functions of coordinate $x$, assuming that $c_{2}(x)$ is
positive-definite, while both the potential $V(x)+iW(x)$ and the
nonlinearity $\hat{N}[\psi ,\psi _{x},|\psi |^{2},(|\psi |^{2})_{x}]$
satisfy the requirement of the $\mathcal{PT}$ symmetry: $V(-x)=V(x)$, $%
W(-x)=-W(x)$, $[\mathcal{PT},\hat{N}]=0$.

\v{\bf Acknowledgments} \\

The work of Y.C. was supported by the NSFC under Grant Nos. 12001246 and
11947087, the NSF of Jiangsu Province of China (No. BK20190991) and the NSF
of Jiangsu Higher Education Institutions of China (No. 19KJB110011). The
work of Z.Y. was supported by the NSFC under Grant Nos. 11925108 and
11731014. The work of B.A.M. was supported, in part by the Israel Science
Foundation though grant No. 1286/17.

\end{document}